\documentclass[prl,amsmath,amssymb,amsfonts,twocolumn,nofootinbib]{revtex4-1}
\usepackage{bm,graphicx,mathrsfs}
\usepackage{graphicx}
\usepackage{epsfig}
\usepackage{amsmath,bbm}
\usepackage{amsfonts,amssymb}
\usepackage{times}

\usepackage{color}
\usepackage{verbatim}

\newcommand{\bR}{\mathbbm{R}}

\newcommand{\cL}{{\cal L}}

\newcommand{\cO}{\mathcal{O}}

\newcommand{\bs}[1]{{\boldsymbol{{#1}}}}
\newcommand{\aalpha}{{\bs{\alpha}}}
\newcommand{\bbeta}{{\bs{\beta}}}
\newcommand{\cnot}{\mathrm{cnot}}
\newcommand{\trace}{{\mathrm{Tr}}}
\newcommand{\calU}{{\cal U }}

\newcommand{\calD}{{\cal D}}

\newcommand{\calA}{{\cal A }}

\newcommand{\calV}{{\cal V }}
\newcommand{\calX}{{\cal X }}
\newcommand{\calY}{{\cal Y }}
\newcommand{\calZ}{{\cal Z }}

\newcommand{\calS}{{\cal S }}
\newcommand{\calI}{{\cal I }}
\newcommand{\calP}{{\cal P }}
\newcommand{\calO}{{\cal O }}

\newcommand{\1}{\mathbbm{1}}

\newcommand{\tr}[1]{{\rm tr}\left({#1}\right)}
\newcommand{\ptr}[2]{{\rm tr}_{#1}\left({#2}\right)}

\newcommand{\raw}{\rightarrow}

\newcommand{\ket}[1]{|#1\rangle}

\newcommand{\proj}[1]{|#1\rangle\langle #1|}
\newcommand{\avr}[1]{\langle#1\rangle}
\newcommand{\half}{\frac{1}{2}}
\def\>{\rangle}
\def\<{\langle}

\newcommand{\Sp}{\,\,\,\,\,\,}
\newcommand{\no}{\nonumber\\}

\newcommand{\be}{\begin{equation}}
\newcommand{\ee}{\end{equation}}
\newcommand{\bq}{\begin{eqnarray}}
\newcommand{\eq}{\end{eqnarray}}
\newcommand{\bea}{\begin{eqnarray}}
\newcommand{\eea}{\end{eqnarray}}

    {\hspace*{\fill}$\Box$\vspace{1.5ex}\par}
    
\newcommand{\qed}{}

\def\qed{\leavevmode\unskip\penalty9999 \hbox{}\nobreak\hfill
     \quad\hbox{\leavevmode  \hbox to.77778em{%
               \hfil\vrule   \vbox to.675em%
               {\hrule width.6em\vfil\hrule}\vrule\hfil}}
     \par\vskip3pt}
     
\begin{document}

\definecolor{james}{rgb}{1,.6,0}
\newcommand{\mjk}[1]{{\color{james} #1}}
\newcommand{\je}[1]{{\color{green} #1}}

\title{Error mitigation for short-depth quantum circuits}

\author{Kristan Temme, Sergey Bravyi and Jay M. Gambetta}
\affiliation{IBM T.J. Watson Research Center, Yorktown Heights NY 10598}
\date{\today}
\begin{abstract}
Two schemes are presented that mitigate the effect of errors and decoherence in short-depth quantum circuits. The size of the circuits for which these techniques can be applied is limited by the rate at which the errors in the computation are introduced. Near-term applications of early quantum devices, such as quantum simulations, rely on accurate estimates of expectation values to become relevant. Decoherence and gate errors lead to wrong estimates of the expectation values of observables used to evaluate the noisy circuit. The two schemes we discuss are deliberately simple and don't require additional qubit resources, so to be as practically relevant in current experiments as possible. The first method, extrapolation to the zero noise limit, subsequently cancels powers of the noise perturbations by an application of Richardson's deferred approach to the limit.  The second method cancels errors by resampling randomized circuits according to a quasi-probability distribution.
\end{abstract}
\maketitle

From the time quantum computation generated wide spread interest, the strongest objection to its viability was the sensitivity to errors and noise. In an early paper, William Unruh \cite{unruh1995maintaining} found that the coupling to the environment sets an ultimate time and size limit for any quantum computation. This initially curbed the hopes that the full advantage of quantum computing could be harnessed, since it set limits on the scalability of any algorithm.  This problem was, at least in theory, remedied with the advent of quantum error correction \cite{shor1995scheme,steane1996error,calderbank1996good}. It was proven that if both the decoherence and the imprecision of gates could be reduced below a finite threshold value, then quantum computation could be performed indefinitely \cite{aharonov1997fault,kitaev1997quantum}.  Although it is the ultimate goal to reach this threshold in an experiment that is scalable to larger sizes, the overhead that is needed to implement a fully fault-tolerant gate set with current codes \cite{fowler2012surface} seems prohibitively large \cite{jones2012layered,devitt2013requirements}. In turn, it is expected that in the near term the progress in quantum experiments will lead to devices with dynamics, which are beyond what can be simulated with a conventional computer. This leads to the question: what computational tasks could be accomplished with only limited, or no error correction?

The suggestions of near-term applications in such quantum devices mostly center around quantum simulations with short-depth circuit \cite{peruzzo2014variational,mcclean2015theory,wecker2015progress} and approximate optimization algorithms \cite{farhi2014quantum}. Furthermore, certain problems in material simulation may be tackled by hybrid quantum-classical algorithms \cite{bauer2015hybrid}. In most such applications, the task can be abstracted to applying a short-depth quantum circuits to some simple initial state and then estimating the expectation value of some observable after the circuit has been applied. This estimation must be accurate enough to achieve a simulation precision comparable or exceeding that of classical algorithms. Yet, although the quantum system evolves  coherently for the most part of the short-depth circuit, the effects of decoherence already become apparent as an error in the estimate of the observable. For the simulation to be of value, the effect of this error needs to be mitigated.  

In this paper we introduce two techniques for {\it quantum error mitigation} that increase the quality of any such short-depth quantum simulations. We find that the accuracy of the expectation value can be increased significantly in the presence of noise. We are looking for error mitigation techniques that are as simple as possible and don't require additional quantum resources. Both techniques require that some noise parameter taken together with system size and circuit depth can be considered a small number.  The first scheme does not make any assumption about the noise model other than it being weak and constant in time. In comparison, the second scheme can tolerate stronger noise; however, it requires detailed knowledge of the noise model. \\ 

{\it Extrapolation to the zero noise limit:} It is our goal to estimate the expectation value of some quantum observable $A$ with respect to an evolved state  $\rho_\lambda(T)$  after time $T$ that is subject to noise characterized by the parameter $\lambda$ in the limit where $\lambda \raw 0$. To achieve this, we apply Richardson's deferred approach to the limit to cancel increasingly higher orders of $\lambda$ \cite{richardson1927deferred}. 

Although gates are typically used to describe quantum circuits, for our analysis it is more convenient to consider the time-dependent Hamiltonian dynamics implementing the circuit. The time-dependent multi-qubit Hamiltonian is denoted by $K(t)$. It can be expanded into $N$ - qubit Pauli operators $P_\alpha \in \avr{\1,X_j,Y_j,Z_j }_{j=1\ldots N}$, where $X_j,Y_j,Z_j$ acts as a single-qubit Pauli matrix on site $j$ and trivially elsewhere. We allow for time-dependent coupling coefficients $J_\alpha(t) \in \bR$. The circuit is encoded as  $ K(t) = \sum_{\alpha} J_\alpha(t) P_\alpha$. The total evolution of the open system with initial state $\rho_0$ will be described by an equation of the following form:
\be
	\frac{\partial}{\partial t} \rho(t) = -i[K(t),\rho(t)] + \lambda \cL(\rho(t))
\ee
for time $t \in [0,T]$. We do not specify the exact form of the generator $\cL(\rho)$ but only require that it is invariant under time rescaling and independent from the parameters $J_\alpha(t)$ in $K(t)$. The noise term $\cL(\rho)$ could be given as a Lindblad operator, or it could correspond to a Hamiltonian that couples to a bath to model non-Markovian dynamics. We ask that there is a parameter $\lambda \ll1$ that indicates a weak action of the noise and that we can bound $\|\cL_{I,t_1} \circ \cL_{I,t_2} \circ \ldots  \circ\cL_{I,t_n} (\rho)\|_1 \leq l_n$, where at most $l_n = \cO(N^n)$. The map $\cL_{I,t}$ is  short-hand notation for the transformation of $\cL$ into the interaction frame generated by $K(t)$. 

The expectation value of the observable $A$ is obtained from the final state $\rho_\lambda(T)$ as $ E_{K}(\lambda) = \tr{A \rho_\lambda(T)} $. The function $E_{K}(\lambda)$ can be expressed as a series in $\lambda$ where the contribution with $\lambda^0$ corresponds to the noise-free evolution.  This can be seen by transforming the evolution into the interaction frame w.r.t $K(t)$ and expanding the Born series, c.f.\ supp.\ mat.\ sec I.  Starting from the noise-free expectation value $E^* = \tr{A \rho_0(T)}$, the expansion is given by
\be\label{expansion}
	E_{K}(\lambda) = E^* + \sum_{k=1}^n a_k \lambda^k + R_{n+1}(\lambda,\cL,T).
\ee
The $a_k$ are model-specific constants typically growing like $a_k \sim N^kT^k$. Here $R_{n+1}(\lambda,\cL,T)$ is the remainder of the expansion and can be bounded by  $ \left |R_{n+1}(\lambda,\cL,T) \right | \leq \|A\| l_{n+1} (\lambda T)^{n+1} / (n+1)!$ by standard arguments. Since we assumed an extensive scaling of $l_n$, such an expansion is only meaningful whenever $ N T \lambda $ is small. We are of course interested in $\lim_{\lambda \raw 0} E_{K}(\lambda) = E^*$; however, we are faced with a small but finite parameter $\lambda$. Since we only have access to $E_{K}(\lambda)$, our estimate of $E^*$ will be off by $\cO(\lambda)$.

This estimate can be improved by Richardson's deferred approach to the limit \cite{richardson1927deferred,sidi2003practical}. To explain the idea, let us assume we can run the quantum circuit at different noise rates $\lambda_j$, with $j = 0,\ldots,n$ and obtain experimental estimates $ \hat{E}_{K}(\lambda_j) = E_{K}(\lambda_j) + \delta_j$. Here the $\lambda_j = c_j \lambda$ are appropriately rescaled values of the experimental noise rate $\lambda$. The estimate deviates from the actual expectation value due to experimental inaccuracies and finite sampling errors by an error $\delta_j$. The estimate of $E^*$ can be significantly improved by considering the approximation $\hat{E}^n_{K}(\lambda)$, which is written as the linear combination 
\be \label{RichardsonCoeff}
\hat{E}^n_{K}(\lambda) = \sum_{j=0}^n \gamma_j \hat{E}_{K}(c_j\lambda).
\ee 
Here we require the coefficients $\gamma_j$ to satisfy the linear system of equations \cite{sidi2003practical}.
\be\label{RichardsonEqn}
	\sum_{l=0}^n \gamma_j = 1 \Sp \mbox{and} \Sp \sum_{j=0}^n \gamma_j \; c^k_j = 0 \Sp \mbox{for} \; k =1\ldots n.
\ee
The linear combination Eq.~(\ref{RichardsonCoeff}) will be an approximation to $E^*$ up to an error of order $\cO(\lambda^{n+1})$. \\

To obtain estimates at different noise rates $\lambda_j$, we use a rescaling trick. We run the same circuit $n+1$ times with rescaled parameters in $K(t)$. We follow the protocol: 
\begin{enumerate}
\item For $j = 0,\ldots,n$ 
\begin{enumerate}
\item choose a rescaling coefficient $c_j > 1$ ($c_0 = 1$) and evolve $\rho_0$ with rescaled Hamiltonian $K^j(t) = \sum_{\alpha} J^j_\alpha(t) P_\alpha$, where
\bq \label{rescaleCoeff}
J^j_\alpha(t) = c_j^{-1} J_\alpha\left(c_j^{-1} t\right),
\eq 
for time  $T_j = c_j T$.
\item Estimate observable $A$ to obtain $\hat{E}_{K}(c_j\lambda)$.
\end{enumerate}
\item Solve equations (\ref{RichardsonEqn}) and compute $\hat{E}^n_{K}(\lambda)$ as in Eq.~(\ref{RichardsonCoeff}).
\end{enumerate}
A rescaling of the equations shows that the state $\rho^j_\lambda(T_j)$, which evolves under $ \dot{\rho}^j_\lambda = -i[K^j(t),\rho^j] + \lambda \cL(\rho^j)$ for time $T_j$, satisfies $\rho^j_\lambda(T_j) = \rho_{c_j\lambda}(T)$, c.f.\ supp.\ mat.\ sec.\ II. Hence the estimates $\hat{E}_{K}(c_j \lambda) =  \tr{A \rho^j_\lambda(T_j)} + \delta_j$ can be obtained from the $n+1$ runs rescaled according to the protocol.\\

{\it If the protocol is performed for $n+1$ steps, the error between the exact expectation value $E^*$ and the estimator $\hat{E}^n_{K}(\lambda)$ can be bounded by
\be\label{lem_richBound}
	|E^* - \hat{E}^n_{K}(\lambda)| \leq \Gamma_n \left( \delta^* +  \|A\| \; \frac{l_{n+1} (\lambda T)^{n+1}}{(n+1)!} \right).
\ee
Here $\Gamma_n = \sum_{j=0}^n |\gamma_j|c_j^{n+1}$ and $\delta^* = \max_j |\delta_j|$ is the largest experimental error.}\\

This follows from repeated application of the triangle inequality, c.f.\ supp.\ mat.\ sec.\ III. The equations (\ref{RichardsonEqn}) can be solved, and one finds that the coefficients $\gamma_j = \prod_{m \neq j} c_m (c_j - c_m)^{-1}$, so that the constant $\Gamma_n$ can be evaluated. In the literature \cite{sidi2003practical}, several choices for progression of $c_j$ are common. The two most frequent series are exponential decrease (Bulirsch - Stoer) and harmonic decay. In our experiments we are actually increasing the noise rate starting from the optimal value, wheres it is common in the numerical literature to improve the small parameter. The result is, of course, the same.\\

{\it Examples :} To demonstrate this method we will consider three numerical examples. In all the examples the time evolution is given by a Hamiltonian $K(t)$ that encodes a control problem. For a single {\it drift step} we evolve with a Hamiltonian $K_R(t) = U_N({\bf \theta}) K_0 U^\dag_N({\bf \theta})$, where the single qubit product unitary  $U_N({\bf \theta}) \in SU(2)^{\otimes N}$ is chosen Haar-random, and the drift Hamiltonian $ K_0 = \sum_{i,j} J_{i,j} X_i Z_j $ is chosen with respect to a random graph and Gaussian distributed couplings $J_{i,j}$.
\begin{figure}[tb]
\begin{center}
\includegraphics[width=0.5\textwidth]{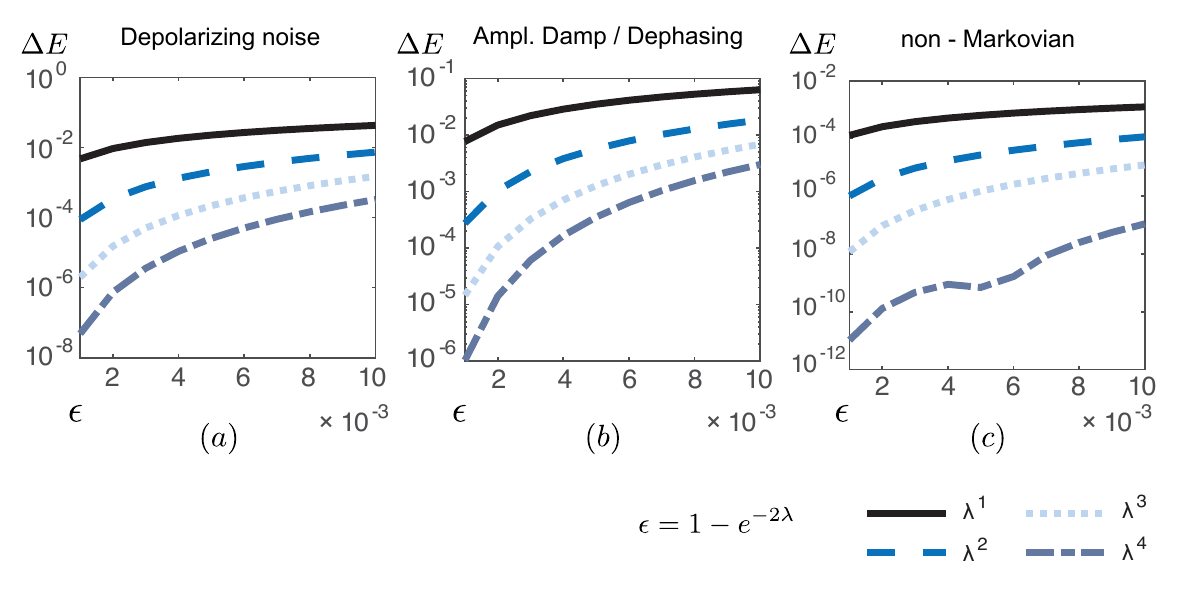}
\end{center}
\vspace*{-4ex}
\caption{(color online) The plots show a random Hamiltonian evolution for $N=4$ system qubits and $d=6$ drift steps, each for time $t = 2$. For all systems plot the error $\Delta E =  |E^* - \hat{E}^n_{K}(\lambda)|$ for $n = 0,1,2,3$. Here $\lambda^{1},n=0$ corresponds to the uncorrected error.  The noise parameter $\lambda = -1/2\log(1-\epsilon)$ is chosen so that all plots have the same perturbation measured in the depolarizing strength $\epsilon = 10^{-3} \ldots 10^{-2}$. The plot shows the mitigation of (a) Depolarizing noise (b) Amplitude damping / dephasing noise and (c) non-Markovian noise, for $\{c_j\}$ chosen as random partition of in the interval $[1,4]$.}
\vspace*{-4ex}
\label{fig:plotRichardson}
\end{figure}
The evolution is subject to three different noise models: first Fig~\ref{fig:plotRichardson}(a), we evolve in the presence of depolarizing noise described by the sum of single qubit generators $\cL_i = -\lambda(2^{-1}\ptr{i}{\rho} - \rho)$ acting on all $N$ qubits. Second, Fig~\ref{fig:plotRichardson}(b), we consider dephasing and amplitude damping noise on every qubit, where we have chosen a ratio of $\lambda_1/\lambda_2 = 1.5$ with a generator $\cL_i = \lambda_1 \left( \sigma_i^- \rho \sigma_i^+ - \half\{\sigma_i^+\sigma_i^-,\rho\}\right) + \lambda_2 \left(Z_i \rho Z_i - \rho \right)$ and $\sigma_i^{\pm} = 2^{-1}(X_i \pm i Y_i)$. Third, Fig~\ref{fig:plotRichardson}(c), we consider a highly non-Markovian setting, where each of the $N$ qubits $i$ is coupled to its own single-qubit bath $b_i$ via the Hamiltonian $V_i = 1/2 \; X_i \otimes X_ {b_i} + 1/2 \; Z_{b_i}$ and the bath is prepared in the initial state $\rho_B = (2\cosh(\beta/2))^{-N}\exp(-\beta \sum_{b_i} \sigma^z_{b_i})$. Then, after the evolution of each noisy circuit $T = td$ we measure a randomly chosen multi-qubit Pauli operator $P_\alpha$.

The graphs in Fig~\ref{fig:plotRichardson} show that with modest effort very high precisions can be obtained. In the low noise range $\epsilon \sim 10^{-3}$ the relative error can be reduced to $\Delta E \sim 10^{-6} - 10^{-11}$. The precision is then essentially determined by the sampling error $\delta^*$, which we have neglected in the plots. \\

{\it Probabilistic error cancellation: } Here we discuss a noise reduction scheme for quantum circuits subject to a Markovian noise. First let us state our assumptions on the noise model. A noisy $N$-qubit device will be described by a basis set of noisy operations $\Omega=\{\calO_1,\ldots,\calO_m\}$ that can be implemented on this device. Each operation $\calO_\alpha$ is a trace-preserving completely positive (TPCP) map on $N$ qubits that acts non-trivially only on a small subset of qubits, say at most two.  For example, $\calO_\alpha$ could be a noisy unitary gate applied to a specified pair of qubits or a noisy qubit initialization. We assume that  noise in the system can be fully characterized such that the map $\calO_\alpha$ is known for each $\alpha$. A  circuit of length $L$ in the basis $\Omega$ is a sequence of $L$ operations from $\Omega$. Let $\Omega_L$ be a set of all length-$L$ circuits in the basis $\Omega$. A circuit $\bs{\alpha}=(\alpha_1,\ldots,\alpha_L)$ implements a map $\calO_{\bs{\alpha}}=\calO_{\alpha_L}\cdots \calO_{\alpha_2} \calO_{\alpha_1}$. The expectation value of an observable $A$ on the final state produced by a noisy circuit ${\bs{\alpha}}$ is 
\[
E({\bs{\alpha}})=\trace{\left[ A\, \calO_{\bs{\alpha}} (|0\rangle\langle 0|^{\otimes n} )\right]}.
\] 
For simplicity, we ignore errors in the initial state  preparation and in the final measurement. Such errors can be accounted for by adding dummy noisy operations before each measurement and after each qubit initialization. Furthermore, we shall assume that $A$ is diagonal in the $Z$-basis  and $\|A\|\le 1$.

Below we show that under certain conditions  the task of simulating an ideal quantum circuit   can be reduced to estimating the expectation value $E({\bs{\alpha}})$  for a suitable random ensemble of noisy quantum circuits $\bs{\alpha}$.  Moreover,  the ideal and the noisy circuits act on the same number of qubits and  have the same depth.

Let $\Gamma=\{\calU_1,\ldots,\calU_k\}$ be a fixed basis set of ideal gates. Each gate $\calU_\beta(\rho) = U_\beta \rho U_\beta^\dag$ is described by a unitary TPCP map on $N$ qubits that acts non-trivially on a small subset of qubits. An ideal length-$L$ circuit in the basis $\Gamma$ is a sequence of $L$ gates from $\Gamma$. A circuit $\bs{\beta}=(\beta_1,\ldots,\beta_L)$ implements a map $\calU_{\bs{\beta}}=\calU_{\beta_L}\cdots  \calU_{\beta_2}\calU_{\beta_1}$. Define an ideal expectation value 
\[
E^*(\bs{\beta}) =\trace{\left[ A\, \calU_{\bs{\beta}} (|0\rangle\langle 0|^{\otimes n} )\right]}.
\]
We consider a simulation task where the goal is to estimate $E^*(\bs{\beta})$ with  a specified precision  $\delta$.

The key idea of our scheme is to represent the ideal circuit as a quasi-probabilistic mixture of noisy ones.  Let us  say that a noisy basis $\Omega$ simulates an ideal circuit $\bs{\beta}$ with the overhead $\gamma_{\bs{\beta}}\ge 1$ if there exists a probability distribution $P_{\bs{\beta}}(\bs{\alpha})$ on the set of noisy circuits $\aalpha\in \Omega_L$ such that
\begin{equation}
\label{ppr1}
\calU_{\bs{\beta}} = \gamma_{\bs{\beta}} \sum_{\bs{\alpha}\in \Omega_L} P_{\bs{\beta}}(\bs{\alpha}) \sigma_{\bs{\beta}}
(\bs{\alpha}) \calO_{\bs{\alpha}}
\end{equation}
for some coefficients  $\sigma_{\bs{\beta}}(\bs{\alpha})=\pm 1$. We also require that the distribution $P_{\bs{\beta}}(\bs{\alpha})$ is sufficiently simple so that one can efficiently sample $\bs{\alpha}$ from $P_{\bs{\beta}}(\bs{\alpha})$. The coefficients $\gamma_{\bbeta},\sigma_{\bs{\beta}}(\bs{\alpha})$ must be efficiently computable. We shall refer to Eq.~(\ref{ppr1}) as a {\em quasi-probability representation} (QPR) of the ideal circuit $\bbeta$. Note that $\gamma_{\bs{\beta}}\ge 1$ because $\calU_{\bs{\beta}}$ and $\calO_{\bs{\alpha}}$ are trace-preserving. Quasi-probability distributions have been previously used to construct classical algorithms for simulation of quantum circuits~\cite{Pashayan2015,Delfosse2015}. Our work can be viewed as an application of these methods to the problem of simulating ideal quantum circuits by noisy ones.

Substituting Eq.~(\ref{ppr1}) into the definition of $E^*(\bs{\beta})$ gives
\begin{equation}
\label{ppr3}
E^*(\bs{\beta})=\gamma_{\bs{\beta}} \sum_{\bs{\alpha}\in \Omega_L} P_{\bs{\beta}} (\bs{\alpha}) \sigma_{\bs{\beta}} (\bs{\alpha}) E({\bs{\alpha}}).
\end{equation}
Let $\aalpha\in \Omega_L$ be a random variable drawn from $P_{\bs{\beta}}(\bs{\alpha})$ and $x\in \{0,1\}^n$ be the final readout  of the noisy circuit $\aalpha$ obtained by measuring each qubit of the final state 
$\calO_{\bs{\alpha}}(|0\rangle\langle 0|^{\otimes n})$ in the $Z$-basis. Note that  $\langle x|A|x\rangle$ is an unbiased estimator of $E({\bs{\alpha}})$ with the variance $O(1)$. Thus from Eq.~(\ref{ppr3}) one infers that  $\gamma_{\bs{\beta}} \sigma_{\bs{\beta}} (\bs{\alpha}) \langle x|A|x\rangle$ is an unbiased estimator of the ideal 
expectation value $E^*(\bs{\beta})$ with the variance $O(\gamma_{\bs{\beta}}^2)$. We can now
estimate $E^*(\bs{\beta})$  with any desired precision $\delta$ by the Monte Carlo method. Define
\begin{equation}
\label{samples}
M= (\delta^{-1} \gamma_{\bs{\beta}})^2
\end{equation}
and generate $M$ samples $\bs{\alpha}^1,\ldots,\bs{\alpha}^M\in \Omega_L$ drawn from $P_{\bs{\beta}}(\bs{\alpha})$. By Hoeffding's inequality, $E^*(\bs{\beta})$ is approximated within error $O(\delta)$ w.h.p. by a random variable 
\begin{equation}
\label{ppr4}
\hat{E}(\bs{\beta})=  \frac{\gamma_{\bs{\beta}}}{M} \sum_{a=1}^M  \sigma_{\bs{\beta}} (\bs{\alpha}^a) \langle x^a|A|x^a\rangle,
\end{equation}
where $x^a\in \{0,1\}^n$ is the final  string of the noisy circuit $\bs{\alpha}^a$. Computing the estimator $\hat{E}(\bs{\beta})$  requires $M$ runs of the noisy circuits, with each run producing a single readout string $x^a$. Estimating  $E^*(\bs{\beta})$ with a precision $\delta$ in the absence of noise by Monte Carlo method would require approximately $\delta^{-2}$ runs. Thus the quantity $\gamma_{\bs{\beta}}^2$ determines the simulation overhead (see Eq.~(\ref{samples})).\\ 
 
A systematic method of constructing QPRs with a small overhead is given in supp.\ mat.\ sec.\ IV. Here we illustrate the method using toy noise models usually studied in the quantum fault-tolerance theory: the depolarizing noise and the amplitude damping noise. For concreteness,  we choose the ideal gate set $\Gamma$ as the standard Clifford+$T$ basis.

\begin{figure}[tb]
\begin{center}
\includegraphics[width=0.38\textwidth]{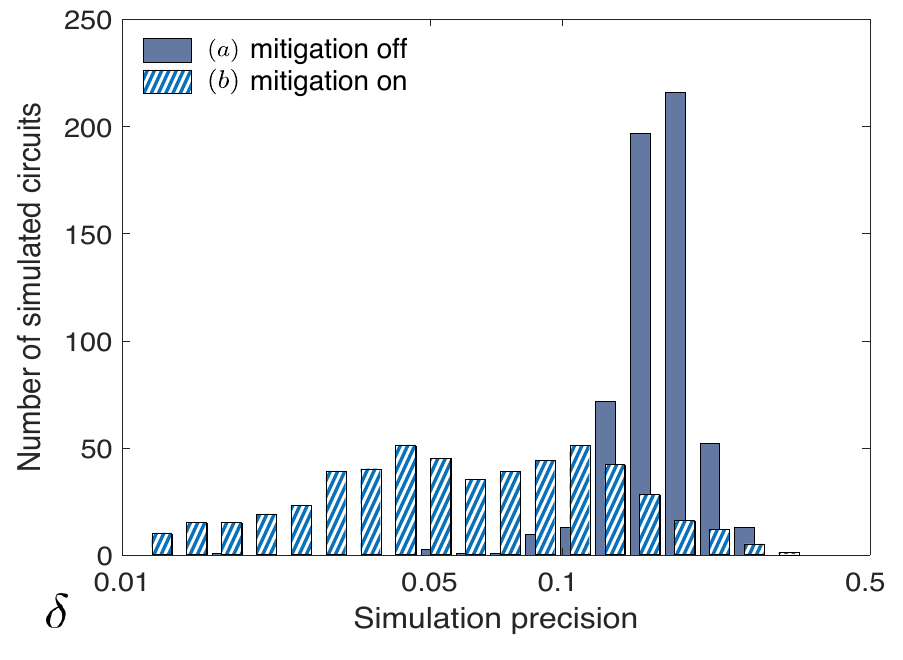}
\end{center}
\vspace*{-4ex}
\caption{Simulation precision $\delta(\beta) = |\hat{E}(\bs{\beta}) - E^*(\bs{\beta})|$ for $500$ randomly generated  ideal Clifford+$T$ circuits on $N=6$ qubits with depth $d=20$. The gates are subject to single- and two-qubit depolarizing noise $\epsilon = 10^{-2}$. The figure shows results for simulations without (a) and with (b) error cancellation. In both cases each ideal circuit was simulated by $M = 4000$ runs of the noisy circuit. For each circuit $\calU_{\bs{\beta}}$ we defined the observable $A$ as a projector  $\Pi_{out}$ onto the subset of $2^{N-1}$ basis vectors with the largest weight in the final state. The results are consistent with  $\gamma_\beta \approx 4.3$ so that $\gamma_\beta M^{-1/2} \approx 0.07$.}
\vspace*{-4ex}
\label{fig:QPRplot}
\end{figure}

Let  $\calD_k$ be the depolarizing noise  on $k = 1,2$ qubits that returns the maximally mixed state with probability $\epsilon$ and does nothing with probability $1-\epsilon$.  Define a noisy version of a $k$-qubit unitary gate $\calU$ as $\calD_k \calU$.  The noisy basis $\Omega$ is obtained by multiplying ideal gates on the left by arbitrary Pauli operators and adding the depolarizing noise. Thus $\Omega$ is a set of operations $\calO_\alpha=\calD_k \calP \calU$, where $\calU\in \Gamma$ is a $k$-qubit ideal gate and $\calP\in \{\calI,\calX,\calY,\calZ\}^{\otimes k}$ is a Pauli TPCP map. The random ensemble of noisy circuits $\calO_{\bs{\alpha}}$  that simulates an ideal circuit $\calU_{\bs{\beta}}$ is constructed in three steps: (1) Start from the ideal circuit, $\calO_{\bs{\alpha}}=\calU_{\bs{\beta}}$.  (2) Modify $\calO_{\bs{\alpha}}$ by adding a Pauli $X,Y,Z$ after each single-qubit gate  with probability $p_1=\epsilon/(4+2\epsilon)$. The gate is unchanged with probability $1-3p_1$. (3) Modify  $\calO_{\bs{\alpha}}$ by adding a  Pauli $IX,IY,\ldots,ZZ$ after each CNOT with probability $p_2=\epsilon/(16+14\epsilon)$. The CNOT is unchanged with probability $1-15p_2$. The resulting circuit is then implemented on a noisy device (which adds the depolarizing noise after each gate) and the final readout string $x$ is recorded. By generating $M$ samples of $x$ one can estimate  $E^*(\bs{\beta})$ from  Eq.~(\ref{ppr4}). The sign function $\sigma_{\bs{\beta}}(\bs{\alpha})$ is equal to $(-1)^r$, where $r$ is the number of Pauli operators added to the ideal circuit $\calU_{\bs{\beta}}$.  As shown in supp. mat. sec. IV,  the above defines a QPR of the ideal circuit $\calU_{\bs{\beta}}$ with the overhead $\gamma_{\bs{\beta}}\approx 1+\epsilon( 3L_1/2 + 15 L_2/8)$, where $L_1$ is the number of single-qubit gates and $L_2$ is the number of CNOTs in the ideal circuit. The method has been tested numerically  for random noisy Clifford+$T$ circuits, see  Fig.~\ref{fig:QPRplot}.

A  more interesting example is the noise described by the amplitude-damping channel $\calA$  that resets every qubit to its ground state with probability $\epsilon$. A noisy version of a $k$-qubit unitary gate $\calU$ is defined as  $\calA^{\otimes k} \calU$. In contrast to the previous example, noisy unitary gates $\calA^{\otimes k} \calU$ alone cannot simulate any ideal unitary gate since $\calA$ is not a unital map. To overcome this, we  extend the noisy basis $\Omega$ by adding noisy versions of single-qubit state preparations $\calA \calP_{\ket{\psi}}$, where  $\calP_{\ket{\psi}}$ maps any input state to $|\psi\rangle \langle \psi|$. Our scheme requires state preparations for single qubit states $\ket{\psi} = \ket{+},\ket{-},\ket{0},\ket{1}$ that can be performed at any time step (not only at the beginning).  In supp.\ mat.\ sec.\ V we show how to construct a QPR of the ideal Clifford+$T$  circuit $\calU_{\bs{\beta}}$ with the overhead $\gamma_{\bs{\beta}}\approx 1+\epsilon( 2L_1 + 4L_2)$. 
The examples considered above suggest that well-characterized noisy circuits can simulate ideal ones with overhead $\gamma\approx (1+c\epsilon)^L$, where  $\epsilon$ is the typical error rate and  $c$  is a  small constant. The value of $c$ can be determined by performing quantum process tomography~\cite{mohseni2008quantum} and finding the QPR for each ideal gate. Using Eq.~(\ref{samples}), one can estimate the number of noisy circuit runs of length $L$ as $M\sim \exp{(2c\epsilon L)}$.  Assuming error rates in the range $\epsilon \sim 10^{-3}$, it may be possible to simulate ideal circuits with $O(10^3)$ gates.\\

{\it Conclusions:} Both error mitigation schemes require no additional quantum hardware such as ancilla or code qubits and work directly with the physical qubits. The zero-noise extrapolation requires sufficient control of the time evolution to implement the rescaled dynamics and hinges on the assumption of a large time-scale separation between the dominant noise and the controlled dynamics. For the probabilistic error cancellation a full characterization of the noisy computational operations is necessary. To obtain this to a precision of $ \sim 10^{-3}$ is challenging in practice. However, if one is willing to sacrifice optimality, a Pauli- or Clifford-twirling~\cite{Silva2008,Magesan2012} can be applied that converts any noise channel into a simple mixture of Pauli errors or depolarizing noise, making the characterization task much more manageable. A very recent independent paper by Li and Benjamin \cite{li2016efficient} discusses similar issues to those addressed here. \\

{\it Acknowledgements:} We thank Antonio Mezzacapo for insightful discussions, and we acknowledge support from the IBM Research Frontiers Institute

%\bibliographystyle{apsrev4-1}
%\bibliography{sample}


%%%%%%%%%%%%%%%%%%%%%    Appendix    %%%%%%%%%%%%%%%%%%%%%%%%
\onecolumngrid
\section{Supplemental Material}
\begin{appendix}

\section{Reducing noise by Richardson extrapolation}

In numerical analysis, Richardson extrapolation \cite{richardson1927deferred,sidi2003practical} is a sequence acceleration method, used to improve the rate of convergence of a sequence.  We use the same technique to extrapolate to the zero-noise limit in short-depth quantum circuits in the presence of noise. We assume that the noise process is constant in time and does not depend on the rescaling of the system Hamiltonian parameters. We consider various noise models in continuous time. 

It is our goal to estimate the expectation value of some observable $A$ with respect to the evolved state $\rho_\lambda(T)$. The actual computation is now encoded in the time-dependent Hamiltonian $K(t)$, and the full evolution is given by the equation
\be
	\frac{\partial}{\partial t} \rho = -i[K(t),\rho] + \lambda \cL(\rho).
\ee
Here we have that 
\be
	K(t) = \sum_{\alpha} J_\alpha(t) P_\alpha
\ee
is some multi-qubit Pauli Hamiltonian with time-dependent coupling constant $J_\alpha(t) \in \bR$ and Pauli operators $P_\alpha \in \avr{\1,X_i,Y_i,Z_i}_{i}$. Starting from some initial state $\rho_0$ the system is evolved for some time $T$. We consider different forms of noise $\cL(\rho)$. As the simplest form of noise, we assume a time-independent Lindblad operator  \cite{lindblad1976generators} of the form 
\be
\cL(\rho) = \sum_{\beta} L_\beta \rho L_\beta^\dag - \frac{1}{2}\{L^\dag_\beta L_\beta,\rho\}.
\ee

However, we can also imagine other forms of errors, such as
\be
\cL(\rho) = -i[V,\rho],
\ee 
where $V$ is some Hamiltonian. This setting in useful when we want to consider more general, possibly non-Markovian noise models, or a noisy evolution derived from first principle \cite{breuer2002theory}. One can make the assumption that the initial state is given by $\rho_0 = \rho_S(0) \otimes \rho_B(0)$ and give the most general form of an interaction Hamiltonian between system and bath, such as 
\be
V = \sum_\alpha S_\alpha \otimes B_\alpha + H_B.
\ee
Here we take the point of view that a small $\lambda$ indicates a separation of time scales, and $\rho_S(0) = \proj{\psi_0}$ may be the initial state of the computation. We assume that $\rho(0)  = \rho_S(0) \otimes \rho_B$, where the bath state is a steady state w.r.t the bath Hamiltonian $[H_B,\rho_B] = 0$. The observable we want to estimate $A = A_S \otimes \1$ is then only supported on the system degrees of freedom.\\

We assume that $\rho_\lambda(T)$ is the state we obtain after the  noisy evolution for time $T$. From this state we can estimate the expectation value of the observable $A$ by various methods. Typically we will sample the expectation value
\be
	E_{K}(\lambda) = \tr{A \rho_\lambda(T)}
\ee
so that an additional sampling error $\delta$ is introduced, and we obtain from out measurement the statistic $\hat{E}_{K}(\lambda) = E_{K}(\lambda) + \delta$. The error can assumed to be asymptotically Gaussian $\delta = \cO\left(M^{-1/2} \sqrt{\tr{\rho_\lambda(T) (A - E_{K}(\lambda))^2}} \right)$ since one typically repeats the experiment $M \gg 1$ times and the i.i.d hypothesis holds.

\subsection{I Series expansion in the noise parameter}

We now show that the function $E_{K}(\lambda)$ can be expressed as a series in $\lambda$ where the contribution with $\lambda^0$ corresponds to the noise-free evolution. We also provide a bound on the error term. To this end, we transform into the interaction picture of $K(t)$. We define $U_K(t) = {\cal T}\left\{\exp(-i \int_0^t K(t') dt')\right\}$, where ${\cal T}\{\cdot\}$ defines the time order expansion. We define the interaction picture through 
\be
	\rho_I(t) = U_K(t) \rho(t) U_K^\dag(t) \Sp \mbox{and} \Sp \cL_{I,t}(\circ) = U_K(t) \cL\left(U_K^\dag(t) \circ U_K(t)\right) U_K^\dag(t),
\ee
where now the generator $\cL_{I,t}$ has become time-dependent. The evolution equation in the interaction picture now reads 
\be
	\partial_t \rho_I(t) = \lambda \cL_{I,t}\left(\rho_I(t)\right).
\ee
Recall that every first-order differential equation can be reformulated as an integral equation
\be
\rho_I(T) = \rho_I(0) + \lambda \int_0^T\cL_{I,t}\left(\rho_I(t)\right) dt.
\ee

This equation can be recursively solved to increasing order in $\lambda$ so that 

\bq
\rho_I(T) = \rho_I(0) + \lambda \int_0^T\cL_{I,t}\left(\rho_I(0)\right) dt  &+&  \lambda^2 \int_0^T\int_0^t \cL_{I,t} \circ \cL_{I,t'} \left(\rho_I(0)\right) dt dt' \no
        &+& \lambda^3 \int_0^T\int_0^t \int_0^{t'} \cL_{I,t} \circ \cL_{I,t'} \circ \cL_{I,t''} \left(\rho_I(0)\right) dt dt' dt''  \ldots.
\eq

Recall that $\rho_I(0) = \rho(0)$. Furthermore, we can conjugate the full expression on both sides with the unitary $U_K(T)$. Let us for notational convenience define $\rho_\lambda(T)$ as the resulting state after evolution with noise rate $\lambda$. We observe that $U_K(T)^\dag \rho(0) U_K(T) = \rho_0(T)$, whereas $U_K(T)^\dag \rho_I(T) U_K(T) = \rho_\lambda(T)$, so that we obtain the expression in the Schr\"odinger picture as 
\bq \label{eqn:expansion_1}
\rho_\lambda(T) = \rho_0(T) &+&  \sum_{k=1}^n \lambda^k  \int_0^T\int_0^{t_1} \ldots \int_0^{t_{k-1}}  U^\dag_K(T) \cL_{I,t_1} \circ \cL_{I,t_2} \circ \ldots \circ\cL_{I,t_k} \left(\rho(0)\right)U_K(T) dt_1 dt_2 \ldots dt_k \no
					     &+&  \lambda^{n+1} \int_0^T\int_0^{t_1} \ldots \int_0^{t_{n}}  U^\dag_K(T) \cL_{I,t_1} \circ \cL_{I,t_2} \circ \ldots \circ\cL_{I,t_n+1} \left(\rho_I(t_{n+1})\right)U_K(T) dt_1 dt_2 \ldots dt_{n+1} .
\eq
The expectation value $E_{K}(\lambda) = \tr{A \rho_\lambda(T)}$ for the observable $A$ can immediately be expanded in a series with parameter $\lambda$ of the form 
\be\label{eqn:expansion_2}
E_{K}(\lambda) = \tr{A \rho_0(T)} + \sum_{k=1}^n a_k \lambda^k + R_{n+1}(\lambda,\cL,T),
\ee
where the constants $a_k$ and the remainder $R_{n+1}(\lambda,\cL,T)$ are obtained by pairing the integrals with the trace $\tr{A \cdot}$ and $ \tr{A \rho_0(T)}  = E^*$ corresponds to the noise-free evolution to which we seek to extrapolate. We read off that 
\be
a_k = \int_0^T\int_0^{t_1} \ldots \int_0^{t_{k-1}}  \tr{U_K(T) A U^\dag_K(T) \cL_{I,t_1} \circ \cL_{I,t_2} \circ \ldots \circ\cL_{I,t_k} \left(\rho(0)\right)} dt_1 dt_2 \ldots dt_k,
\ee
as well as 
\be
R_{n+1}(\lambda,\cL,T) = \lambda^{n+1} \int_0^T\int_0^{t_1} \ldots \int_0^{t_{n}}  \tr{U_K(T) A U^\dag_K(T)  \cL_{I,t_1} \circ \ldots \circ\cL_{I,t_n+1} \left(\rho_I(t_{n+1})\right)} dt_1 \ldots dt_{n+1}.
\ee
We can bound $|R_{n+1}(\lambda,\cL,T)|$ by a simple application of Cauchy's mean value theorem and H\"older's inequality. We observe by first applying the midpoint Theorem that there exist $\xi_1,\ldots,\xi_{n+1}$ so that
\be
R_{n+1}(\lambda,\cL,T) = \frac{\lambda^{n+1}T^{n+1}}{(n+1)!} \tr{U_K(T) A U^\dag_K(T)  \cL_{I,\xi_1} \circ \ldots \circ\cL_{I,\xi_n+1} \left(\rho_I(\xi_{n+1})\right)}.
\ee
We can then of course immediately bound the inner product  
\be
|\tr{U_K(T) A U^\dag_K(T)  \cL_{I,t_1} \circ \ldots \circ\cL_{I,t_n+1} \left(\rho_I(t_{n+1})\right)}| \leq \|A\| \| \cL_{I,t\xi_1} \circ \ldots \circ \cL_{I,\xi_n+1}\left(\rho_I(\xi_{n+1})\right) \|_1
\ee
by a direct application of H\"older's inequality. Note that all Schatten norms are unitarily invariant, so when the map $\cL$ is bounded, we can apply the subsequent operator norm inequalities 
\be
	\| \cL_{I,t\xi_1} \circ \ldots \circ \cL_{I,\xi_n+1}\left(\rho_I(\xi_{n+1})\right) \|_1 \leq \| \cL \|^{n+1}_{1\raw 1}.
\ee 
It is safe to assume that a Lindblad operator $\cL$ acting on a finite dimensional system, such as a collection of qubits, is bounded. However, we also consider the case of a first-principle noise model that can even be non-Markvoian. In this setting the operator $\cL(\rho) = -[V,\rho]$ is expected to couple to an arbitrary large bath and $V$ may contain unbounded operators, such as bosonic operators. In such a setting an upper bound in terms of an operator norm of $\cL$ is a moot point. Yet, in this case we can transform the evolution into the Heisenberg picture $\cL^*$, for the observable $A(0)= A_S \otimes \1 $, and look at the equations for $A(t)$ instead. The almost identical analysis as performed above can be carried through, but this time we can obtain a bound on 
\be
|\tr{\rho_I(0) \cL^*_{I,\xi_{n+1}}\circ \ldots \circ  \cL^*_{I,t\xi_1}(A(t))} | \leq \|A(t)\| \| \cL_{I,t\xi_1} \circ \ldots \circ \cL_{I,\xi_{n+1}}\left(\rho_I(0)\right) \|_1.
\ee
We obtain almost the same type of bound from H\"older's inequality since $\|A(t)\| \leq \|A\|$ for contractive evolutions,  where the sole difference is now that $\| \cL_{I,t\xi_1} \circ \ldots \circ \cL_{I,\xi_{n+1}}\left(\rho_I(0)\right) \|_1 \leq l_{n+1}$ only depends on the initial state. Since we now consider the action of the operators in $V$ on a well-behaved initial state $\rho(0)$, we can assume that $l_{n+1}$ is a reasonable bound. In either case, we will now write for the bound on $|R_{n+1}(\lambda,\cL,T)|$ from now on:
\be
|R_{n+1}(\lambda,\cL,T)| \leq \|A\| \; l_{n+1}\; \frac{\lambda^{n+1} T^{n+1}}{(n+1)!}.
\ee 
The coefficients $a_k$ can be bounded in a similar fashion. Note that, if we assume that noise acts locally on each qubit, such as for instance, when the dissipator $\cL$ corresponds to single-qubit depolarizing noise, so that $\cL(\rho) = \sum_{i=1}^N (\frac{1}{2}\ptr{[i]}{\rho} - \rho)$. We have that $|| \cL ||_{1\raw1} = \cO(N)$ is extensive in the system size. A similar argument holds for the case when the individual qubits couple to a bath. From this we can deduce that for local noise we typically find $l_{k} = \cO(N^{k})$ as mentioned in the main text, and that $|a_k| \leq \cO((N\; T)^k)$. \\

It is also worthwhile to point out the following observation. For different types of error terms $\cL$ it may happen that not all powers of $\lambda$ are present in the expansion. It is conceivable that some system bath interactions could lead to an expansion in only even powers of $\lambda$. If this occurs, the Richardson extrapolation method is particularly efficient,  since a higher order of precision can be obtained with fewer values of $\lambda_j$.

\subsection{II Experimental rescaling of the noise parameter}

In order to apply Richardson extrapolation, we have to be able to evaluate $E_{K}(\lambda)$ for different values of $\lambda$. In an actual experiment, we can't directly control the parameter $\lambda$: however, we may control the evolution $K(t)$. To this end, we introduce a rescaling. We redefine 
\be
\label{rescaling}
	T \raw T' = cT \Sp \mbox{as well as} \Sp J_\alpha(t) \raw J'_\alpha(t) =  c^{-1}J_\alpha(c^{-1} t) \Sp \mbox{from which also} \Sp \rho(t) \raw \rho'(t) = \rho(c^{-1} t).
\ee
We claim that this rescaling maps $\rho'_\lambda(T') = \rho_{c \lambda}(T)$ if the noise operator $\cL$ does not depend on the Hamiltonian couplings $J_\alpha(t)$ and is constant in time. This rescaled density matrix then leads to a new evaluation $E'_{K}(\lambda) \raw E_{K}(c \lambda) $ of the expectation value. To see that the rescaling has the desired effect, we again make use of the integral representation of $\rho_\lambda(T)$, for which we can write now in the Schr\"odinger picture 
\be
\rho_\lambda(T) = \rho(0) - i \int_0^T[K(t) , \rho(t)] dt + \lambda \int_0^T\cL(\rho(t)) dt. 
\ee
We can now choose a re-parametrization of the evolution $c^{-1}J_\alpha(c^{-1} t)$ and an increased runtime $cT$, and write 
\be
\rho'_\lambda(T') = \rho(0) - i \int_0^{cT} [K'(t) , \rho'(t)] dt + \lambda \int_0^{cT} \cL(\rho'(t)) dt 
\ee
with $K'(t) = \sum_\alpha c^{-1}J_\alpha(c^{-1} t)P_\alpha$. If we now substitute the integration variable according to $t = c t'$, we have that $dt = c dt'$, which leads to
\bq\label{rescaleCoeff}
\rho'_\lambda(T') &=& \rho(0) - i \int_0^{T} \sum_\alpha c^{-1}J_\alpha(t')[P_\alpha , \rho(t')] c dt' + \lambda \int_0^T\cL(\rho'(t)) c dt' \no
			   &=&  \rho(0) - i \int_0^{T} \sum_\alpha [K(t'), \rho(t')] dt' + \lambda c \int_0^T\cL(\rho(t')) dt' \no
&=& \rho_{c \lambda}(T).
\eq 
Hence, rescaling the evolution according to equation (\ref{rescaling}) leads to an effective rescaling of the dissipative rate $\lambda$. This can be done for any constant dissipator $\cL$ and allows the experimenter to evaluate $E_{K}(\lambda)$ for different values of $c\lambda$ so that we can apply the Richardson extrapolation procedure.\\

Note that for different experimental circumstances, other rescaling methods of the parameter $\lambda$ may actually be easier to implement. For example, in an optical experiment that is plagued by photon loss, it may be suitable to consider different methods of directly changing the photon loss rate. The only requirement is that the modification of $\lambda_j$ can be performed sufficiently accurately 
so the extrapolation can be performed. 

\subsection{III Error bounds on the noise-free estimator}

Let us now show that the protocol leads to the desired error bound on the estimated expectation value as claimed in the main text. Recall that we first choose a set of $n+1$ rescaling parameters $c_0 = 1 < c_1 < \ldots < c_{n}$, to evolve with respect to the rescaled Hamiltonian $K^j(t)$ for time $T_j = c_j T$. As discussed in the previous section, this evolution leads to a state  $\rho^j_\lambda(T_j) = \rho_{c_j\lambda}(T)$, c.f. Eq.~(\ref{rescaleCoeff}) as was discussed in section II. If we now measure the observable $A$ on these states we obtain for $j = 0 \ldots n+1$ the estimates $ \hat{E}_{K}(c_j \lambda) = E_{K}(c_j \lambda) + \delta_j$. Recall the set of equations for $\gamma_j$ defined in \cite{schneider1975vereinfachte} and given in the main text, which requires for the $\{c_j\}$ that
\bq
	&&\sum_{l=0}^n \gamma_j = 1 \no  
	&&\sum_{j=0}^n \gamma_j \; c^k_j = 0 \Sp \mbox{for} \; k =1\ldots n.
\eq
Now we observe that estimators $\hat{E}_{K}(c_j \lambda)$ can be expressed as  

\be
	\hat{E}_{K}(c_j \lambda) = E^* + \sum_{k=1}^n a_k c_j^k \lambda^k + R(c_j \lambda, \cL, T) + \delta_j
\ee

due to the expansion (\ref{eqn:expansion_2}) discussed in section I. Recall now the definition of our improved estimator $\hat{E}^n_K(\lambda)$ as given in the main text,
$ \hat{E}^n_{K}(\lambda) = \sum_{j=0}^n \gamma_j \hat{E}_{K}(c_j\lambda)$, for which then
\bq
\hat{E}^n_{K}(\lambda) &=& \sum_{j=0}^n \left(\gamma_j E^* + \sum_{k=1}^n a_k c_j^k \lambda^k + R(c_j \lambda, \cL, T) + \delta_j\right) \no
&=&  E^* \left(\sum_{j=0}^n \gamma_j\right)   + \sum_{k = 1}^n  a_k \lambda^k \left( \sum_{j=0}^n \gamma_j c_j^k\right)  +   \left(\sum_{j=0}^n \gamma_j R(c_j \lambda, \cL, T) + \delta_j\right).
\eq
Recall the equations for $\gamma_j$ from which we can then infer after the application of the triangle inequality
\be
| E^* - \hat{E}^n_{K}(\lambda)| \leq \sum_{j=0}^n |\gamma_j| \left( |R(c_j \lambda, \cL, T)| + |\delta_j| \right).
\ee
After the application of the bound $|R(c_j \lambda, \cL, T)| \leq  \|A\| \; l_{n+1}\;  c_j^{n+1} \lambda^{n+1} T^{n+1} ((n+1)!)^{-1}$ and the observation that $c_j \geq 1$, 
we can bound the difference with $\Gamma_n = \sum_{j=0}^n |\gamma_j| c_j^{n+1}$ and obtain the final bound
\be
| E^* - \hat{E}^n_{K}(\lambda)| \leq \Gamma_n \left( \delta^* +  \|A\| \frac{l_{n+1} \lambda^{n+1} T^{n+1}}{(n+1)!}\right), 
\ee 
with $\delta^* = \max_j |\delta_j|$.

In the Richardson extrapolation literature \cite{sidi2003practical}, two types of sequences $c_j$ are considered frequently.  In the Bulirsch - Stoer series the rescalings are chosen so that $c_j = h^j c_0$ constitutes an exponential series, which is typically chosen at base $h = 1/2$; but harmonic series have also been frequently applied, e.g. for parameters $q > 1, \eta \geq 0$ one can choose $c_j = (j + \eta)^{-q} c_0$. Note that in our experiments we are actually increasing the noise rate starting from the optimal value, whereas it is common in the numerical literature to improve the small parameter, so that $c_{j+1} \leq c_j$.  The result here is of course the same, and just corresponds to a reordering of the labels when $n$ is finite. For both of the aforementioned cases, a bound on $\Gamma_n$ has been derived. One is mostly interested in the asymptotic behavior of $\Gamma_n$ as $n \raw \infty$ in order to analyze the numerical stability of the method. In current experiments we only expect to go to third or forth order, making the stability analysis less relevant.

\section{Probabilistic error cancellation by resampling}
The key idea of our scheme is to represent the ideal circuit as a quasi-probabilistic mixture of noisy ones. Central to this approach is the quasi probability representation (QPR) of the noise-free circuit $\calU_{\bs{\beta}}$.  We note that quasi-probability distributions have been previously used to construct classical algorithms for simulation of quantum circuits~\cite{Pashayan2015, Delfosse2015}. Our work can be viewed as an application of these methods to the problem of simulating ideal quantum circuits by noisy ones. \\

The general approach to constructing a QPR for a quantum circuit is the following:  Suppose you are given a set of noisy operations  $\Omega=\{\calO_1,\ldots,\calO_m\}$ that can be implemented on a noisy $N$-qubit device. We assume that we can perform gate tomography \cite{mohseni2008quantum} to specify the gates with an accuracy that is comparable to the desired accuracy of the ideal circuit. These noisy operations are noisy versions of ideal quantum gates and are assumed to form a full basis of TPCP operations, i.e. an element $\cO_k \in \Omega$ is always of the form
\be
\calO_k(\rho) = \sum_i O_{i,k} \rho O_{i,k}^\dagger \Sp  \mbox{with} \Sp \sum_i O_{i,k}^\dagger O_{i,k} = \1.
\ee
A crucial condition is that the set of noisy operators $\Omega$ constitutes a basis in the space of TPCP operations that is sufficiently large, so that any ideal, unitary gate $\calU(\rho) = U\rho U^\dagger$ can be expressed as a linear combination of noisy gates in $\Omega$. Hence, there have to be coefficients $\eta_\alpha \in \bR$, and noisy operations in $\calO_\alpha \in \Omega$ so that we can write for any ideal gate in the circuit
\be
	\calU(\rho) = \sum_\alpha \eta_\alpha \calO_\alpha(\rho), \Sp \forall \rho.
\ee
This linear expansion of $\calU$ can then be cast into the form of a quasi-probability representation.\\

\begin{figure}[h]
\begin{center}
\includegraphics[width=0.70\textwidth]{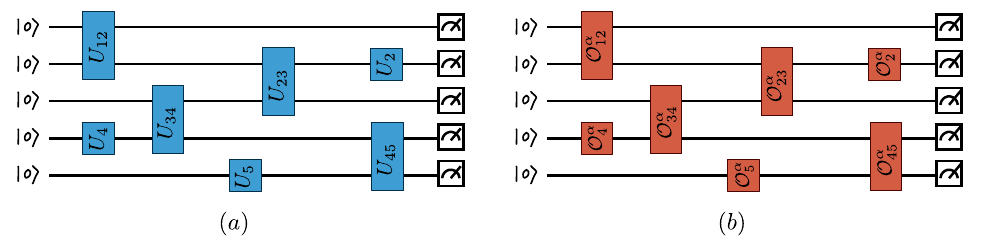}
\end{center}
\vspace*{-3ex}
\caption{(color online) The figure (a) represents the ideal circuit we want to simulate. It is comprised of single- and two-qubit gates $\{U_{12},\ldots,U_{5} \}$. We assume that a complete set of noisy gates exist $\Omega=\{\calO_{12}^{\alpha_{12}},\ldots,\calO_{5}^{\alpha_{5}}\}$, which serve as an operator basis in which the action of the ideal set can be expanded. It is then sufficient to sample circuits, as given in figure (b), where the gates are drawn from the probability distribution $P_{\bs{\beta}}$ in Eq.~(\ref{ppr1}). }
\label{fig:sample_Rand}
\end{figure}

On real quantum devices we can only apply the noisy operations $\Omega$. We say that a circuit of length $L$ in the basis $\Omega$ is a sequence of $L$ operations from $\Omega$. Such a circuit, c.f.\ Fig~\ref{fig:sample_Rand}(b), indexed by $\bs{\alpha}=(\alpha_1,\ldots,\alpha_L)$ implements a noisy map $\calO_{\bs{\alpha}}=\calO_{\alpha_L}\cdots \calO_{\alpha_2} \calO_{\alpha_1}$. 

The expectation value of an observable $A$ on the final state produced by a noisy circuit ${\bs{\alpha}}$ is 
\[
E({\bs{\alpha}})=\trace{\left[ A\, \calO_{\bs{\alpha}} (|0\rangle\langle 0|^{\otimes n} )\right]}.
\] 
For simplicity, we ignore errors in the initial state  preparation and in the final measurement. Such errors can be accounted for by adding dummy noisy operations before each measurement and after each qubit initialization. Furthermore, we shall assume that $A$ is diagonal in the $Z$-basis and $\|A\|\le 1$.

The task of simulating an ideal quantum circuit $\calU_{\bs{\beta}}$, c.f. Fig.~\ref{fig:sample_Rand}(a), of ideal gates $\calU_{\bs{\beta}} = \calU_{\beta_L} \ldots \calU_{\alpha_2} \calU_{\alpha_1}$, can be reduced to estimating the expectation values $E({\bs{\alpha}})$ for a suitable random ensemble of noisy quantum circuits $\bs{\alpha}$. That is, we can obtain estimates for the ideal expectation values  
\[
E^*(\bs{\beta}) =\trace{\left[ A\, \calU_{\bs{\beta}} (|0\rangle\langle 0|^{\otimes n} )\right]},
\]
after the application of the circuit $\calU_{\bs{\beta}}$ by estimating noisy circuit outputs. Moreover, the ideal and the noisy circuits  act on the same number of qubits and have the same length.\\

We  say that the noisy basis $\Omega$ simulates an ideal circuit $\bs{\beta}$ if there exists a probability distribution $P_{\bs{\beta}}(\bs{\alpha})$ on the set of noisy circuits $\aalpha\in \Omega_L$ such that
\begin{equation}
\label{ppr1}
\calU_{\bs{\beta}} = \gamma_{\bs{\beta}} \sum_{\bs{\alpha}\in \Omega_L} P_{\bs{\beta}}(\bs{\alpha}) \sigma_{\bs{\beta}}(\bs{\alpha}) \calO_{\bs{\alpha}}
\end{equation}
for some coefficients  $\sigma_{\bs{\beta}}(\bs{\alpha})=\pm 1$. We require that the distribution $P_{\bs{\beta}}(\bs{\alpha})$ is sufficiently simple so that one can efficiently sample $\bs{\alpha}$ from $P_{\bs{\beta}}(\bs{\alpha})$. The coefficients $\gamma_{\bbeta},\sigma_{\bs{\beta}}(\bs{\alpha})$ must be efficiently computable. \\

We can see that the estimates of the noisy circuit are related to the ideal circuit probability by substituting Eq.~(\ref{ppr1}) into the definition of $E^*(\bs{\beta})$. This gives
\begin{equation}
\label{ppr3}
E^*(\bs{\beta})=\gamma_{\bs{\beta}} \sum_{\bs{\alpha}\in \Omega_L} P_{\bs{\beta}} (\bs{\alpha}) \sigma_{\bs{\beta}} (\bs{\alpha}) E({\bs{\alpha}}).
\end{equation}
The construction of QPRs with optimal overhead of general operation-dependent noise is an interesting  open problem. A preliminary analysis shows that a noisy basis $\Omega$ that includes  noisy versions of all single-qubit and two-qubit Clifford gates, $T$-gates, and noisy qubit initializations  in the $X,Y,Z$ basis can simulate any ideal gate $\calU_\beta$ from the Clifford+$T$ gate set with the overhead $\gamma_\beta\le 1+O(\epsilon)$, provided that each noisy operation is $\epsilon$-close to its ideal analogue. Unfortunately,  the constant coefficient in this upper bound is far too large to have any practical implications. 

Furthermore,  the full Clifford group on two-qubits contains $11520$ gates. It may not be feasible to perform process tomography for each of those gates. We shall see that for certain noise models, such as the amplitude damping noise, QPRs can be constructed only if the noisy basis $\Omega$ includes some qubit initialization maps. In particular, noisy circuits $\aalpha$ that appear in Eq.~(\ref{ppr1}) may apply qubit initializations at intermediate time steps, even though the ideal circuit $\bbeta$ initializes all the qubits at the very first step. All QPRs  constructed below preserve the circuit depth. That is, if the ideal circuit $\bbeta$ has depth $d$ then all noisy circuits $\aalpha$ in Eq.~(\ref{ppr1})  have depth at most $d$.

\subsection{IV Minimal overhead decomposition of noise free circuit}
Let us discuss how to construct QPRs with a small overhead. For concreteness,  we choose the ideal gate set $\Gamma$ as the Clifford+$T$ basis. It includes the identity gate $I$, the Hadamard gate $H$, phase-shift gates $S=\mathrm{diag}[1,i]$ and $T=\mathrm{diag}[1,e^{i\pi/4}]$, and the CNOT. For technical reasons, we shall assume that each CNOT is followed by single-qubit gates (that could be identity gates). We shall consider toy noise models  usually studied in the quantum fault-tolerance theory: the depolarizing noise and the amplitude damping noise. 

First let us describe product QPRs  that can be constructed independently for each gate in the ideal circuit. Consider a fixed ideal gate $\calU_\beta\in \Gamma$. Let $\calO_1,\ldots,\calO_p\in \Omega$ be the list of all noisy operations  whose support is contained in the support of $\calU_\beta$. Consider the following linear program with $2p$ real variables $\mu_1,\ldots,\mu_p,\eta_1,\ldots,\eta_p$.
\begin{equation}
\label{LP}
\mbox{\bf minimize}  \quad  \sum_{\alpha=1}^p \mu_\alpha
\end{equation} 
\begin{equation}
\label{LP1}
\mbox{\bf subject to} \quad \left\{ \begin{array}{rcl}
\eta_\alpha &\le&  \mu_\alpha \\
-\eta_\alpha &\le & \mu_\alpha \\
\calU_{\beta}&=& \sum_{\alpha=1}^p  \eta_\alpha \calO_\alpha. \\
\end{array} \right.
\end{equation}
Suppose $\{\mu_\alpha,\eta_\alpha\}$ is the optimal solution of the program. Note that $\mu_\alpha=|\eta_\alpha|$ for all $\alpha$ since otherwise the objective function can be decreased. 
Define $\gamma_\beta=\sum_{\alpha=1}^p \mu_p$, $P_\beta(\alpha)=\mu_\alpha/\gamma_\beta$, and $\sigma_\beta(\alpha)=\mathrm{sgn}{(\eta_\alpha)}$. Then
\begin{equation}
\label{local}
\calU_\beta =\gamma_\beta \sum_{\alpha=1}^p P_\beta(\alpha) \sigma_\beta(\alpha) \calO_\alpha,
\end{equation}
which is a gate-wise version of the  QPR Eq.~(\ref{ppr1}). We shall say that a noisy basis $\Omega$ simulates a gate $\calU_\beta$ with the overhead $\gamma_\beta$ if the linear program Eqs.~(\ref{LP},\ref{LP1}) has a feasible solution with value $\gamma_\beta$. A  product QPR of the ideal circuit $\bbeta$ is defined as a product of all gate-wise QPRs Eq.~(\ref{local}). It gives  $\gamma_{\bs{\beta}}=\gamma_{\beta_1}\cdots \gamma_{\beta_L}$,  $P_{\bs{\beta}} (\bs{\alpha})= P_{\beta_1}(\alpha_1) \cdots P_{\beta_L}(\alpha_L)$ and  $\sigma_{\bs{\beta}} (\bs{\alpha})=\sigma_{\beta_1}(\alpha_1) \cdots \sigma_{\beta_L}(\alpha_L)$. The assumption that all noisy operations $\calO_\alpha$ in Eq.~(\ref{LP1}) act non-trivially only within the support of $\calU_\beta$ allows one to restrict Eq.~(\ref{LP1}) to operations acting on at most two qubits. Such operations can be represented by real matrices of size $16\times 16$ by computing matrix elements of $\calO_\alpha$ and $\calU_\beta$ in the Pauli basis. Thus the program Eqs.~(\ref{LP},\ref{LP1}) can be solved in time $O(1)$.  Since the ideal gate set has size $O(n^2)$, product QPRs can be  computed in time $O(n^2)$. Furthermore, if two ideal gates have disjoint supports, then the gate-wise QPRs defined in Eq.~(\ref{local}) have disjoint supports. Thus product QPRs preserve the circuit depth.

\subsection{V Depolarizing noise cancellation and numerical results}
Let us illustrate the construction of  product QPRs using the depolarizing noise as an example. Let $\calD_k$ be the $\epsilon$-depolarizing channel on $k$ qubits that returns the maximally mixed state with probability $\epsilon$ and does nothing with probability $1-\epsilon$.  Define a noisy version of a $k$-qubit unitary gate $\calU$ as $\calD_k \calU$. Define a noisy basis $\Omega$ by multiplying ideal gates on the left by arbitrary Pauli operators and adding the depolarizing noise. Thus $\Omega$ is a set of operations $\calO_\alpha=\calD_k \calP \calU$, where $\calU\in \Gamma$ is a $k$-qubit ideal gate and $\calP\in \{\calI,\calX,\calY,\calZ\}^{\otimes k}$ is a Pauli TPCP map. A Pauli map $\calP$ corresponding to a Pauli operator $P\in \{I,X,Y,Z\}$ is defined by  $\calP(\rho)=P\rho P$. Here $k=1,2$. We claim that $\Omega$ simulates ideal single-qubit  gates $\calU_\beta\in \Gamma$  with  the overhead $\gamma_\beta=(1+\epsilon/2)/(1-\epsilon)$ and simulates CNOTs with the overhead $\gamma_\beta=(1+7\epsilon/8)/(1-\epsilon)$. 

Indeed, suppose $\calU_\beta\in \Gamma$ is a single-qubit gate. Let us look for a  solution of Eq.~(\ref{LP1}) in the form $\calO_\alpha=\calD_1 \calP \calU_\beta$, where $\calP\in \{\calI,\calX,\calY,\calZ\}$.
Then Eq.~(\ref{LP1}) is equivalent to  

\[
\calD_1^{-1}=\eta_1 \calI + \eta_2 \calX + \eta_3 \calY + \eta_4\calZ.
\] 

One can easily check that the optimal solution minimizing $\sum_\alpha |\eta_\alpha|$ is $\eta_1=1+3\epsilon/4(1-\epsilon)$ and $\eta_\alpha=-\epsilon/4(1-\epsilon)$ for $\alpha=2,3,4$. Therefore $\gamma_\beta=\sum_\alpha |\eta_\alpha|=(1+\epsilon/2)/(1-\epsilon)$. The CNOT  is simulated in a similar fashion by representing $\calD_2^{-1}$ as a linear combination of  two-qubit Pauli maps. 
The random ensemble of noisy circuits  $\calO_{\bs{\alpha}}$ that simulates an ideal circuit $\calU_{\bs{\beta}}$ is constructed in three steps:

\begin{enumerate}
\item Start from the ideal circuit, $\calO_{\bs{\alpha}}=\calU_{\bs{\beta}}$. 
\item Modify $\calO_{\bs{\alpha}}$ by adding a Pauli $X,Y,Z$ after each single-qubit gate  with probability $p_1=\epsilon/(4+2\epsilon)$. The gate is unchanged with probability $1-3p_1$.
\item Modify  $\calO_{\bs{\alpha}}$ by adding a  Pauli $IX,IY,\ldots,ZZ$ after each CNOT with probability $p_2=\epsilon/(16+14\epsilon)$. The CNOT is unchanged with probability $1-15p_2$. 
\end{enumerate}

The resulting circuit is then implemented on a noisy device (which adds the depolarizing noise after each gate) and the final readout string $x$ is recorded. By generating $M$ samples of $x$ one can estimate 
$E^*(\bs{\beta})$  using  Eq.~(10) of the main text. The sign function $\sigma_{\bs{\beta}}(\bs{\alpha})$ is equal to $(-1)^r$, where $r$ is the number of Pauli operators added to the ideal circuit $\calU_{\bs{\beta}}$ to obtain $\calO_{\bs{\alpha}}$.

\subsubsection{Numerical simulations}
The error cancellation method was tested numerically for small Clifford+$T$ circuits subject to the depolarizing noise.  We choose the ideal circuit $\calU_{\bs{\beta}}$ as a composition of $d$ alternating layers of gates, with each layer being  either a tensor product of $n$ single-qubit gates $I,H,S,T$ (for odd layers) or a tensor product of $n/2$ CNOTs (for even layers).  The resulting circuit $\calU_{\bs{\beta}}$ has depth $d$.  Simulations were performed for $500$ random circuits $\calU_{\bs{\beta}}$ as above with the initial state $|+\rangle^{\otimes n}$. Each single-qubit gate was picked randomly from the set $\{I,H,S,T\}$. Control and target qubits for each CNOT were picked at random. 

\begin{figure*}[ht]
\centerline{\includegraphics[height=6cm]{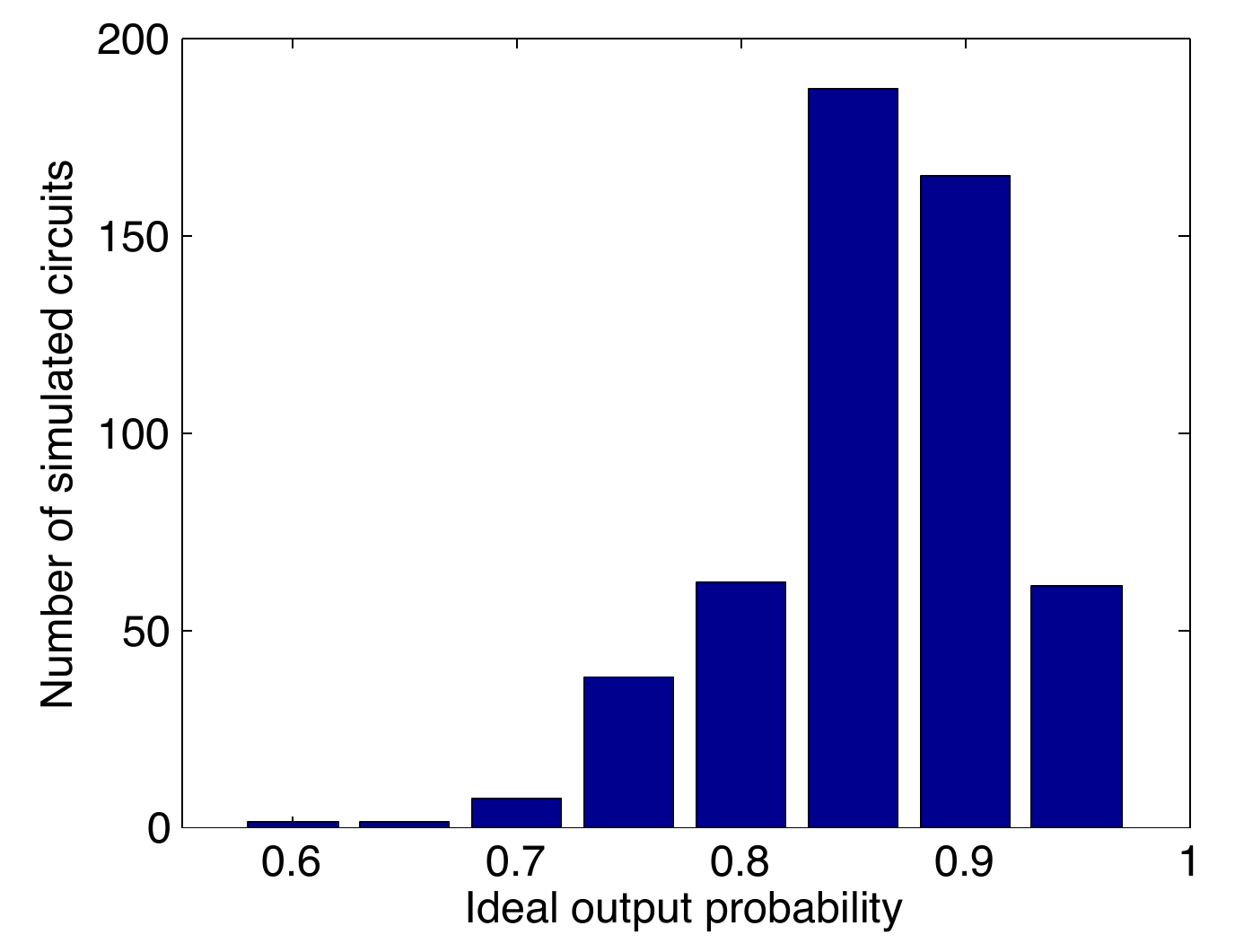}}
\caption{Distribution of the ideal circuits according to their output probability $E^*(\bs{\beta})$. }
\label{fig:plotIdeal}
\end{figure*}

\begin{figure*}[ht]
\centerline{\includegraphics[height=6cm]{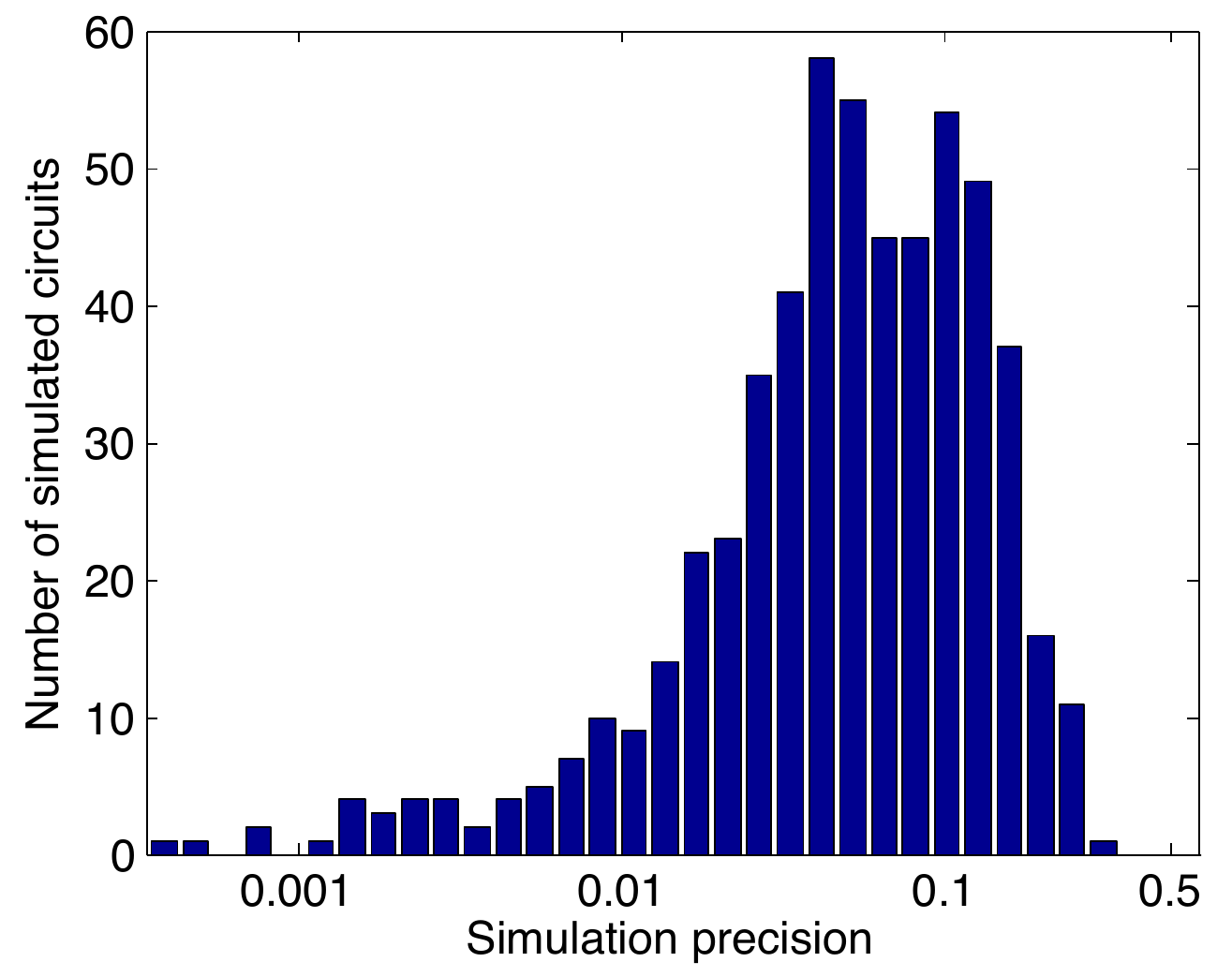}$\quad \quad $ \includegraphics[height=6cm]{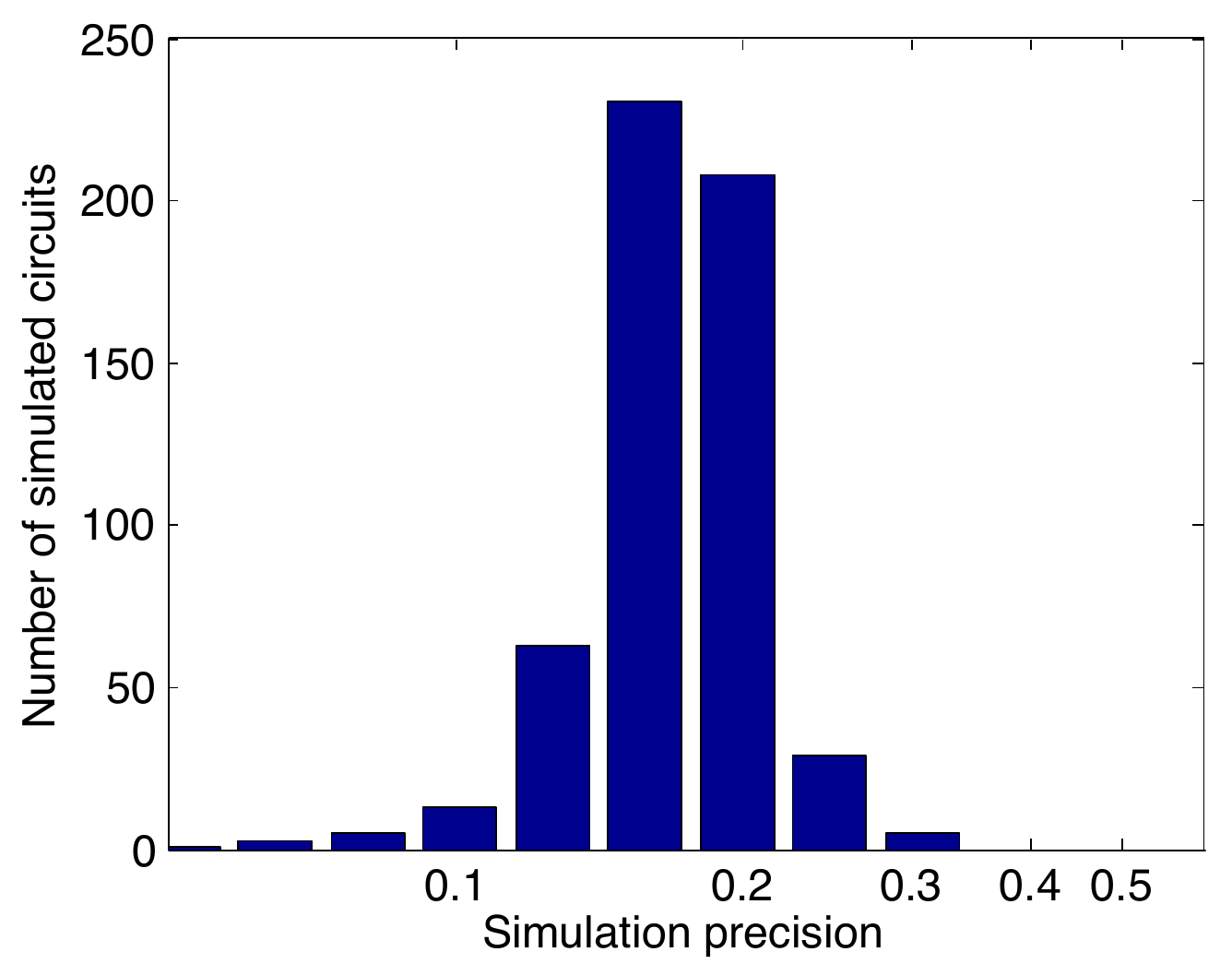}}
\caption{Simulation precision for $\approx 500$ randomly generated  ideal Clifford+$T$ circuits
on $n=6$ qubits with depth $d=20$.  The left and the right panels show results for simulations 
with and without error cancellation. In both cases each ideal circuit was simulated
by $M=4000$ runs of the noisy circuit. }
\label{fig:quasi}
\end{figure*}

For each ideal circuit $\calU_{\bs{\beta}}$ we choose the observable $A$ as a  projector  onto the  subset of $2^{n-1}$ basis states $x\in \{0,1\}^n$  whose probability in the final state of  $\calU_{\bs{\beta}}$ is above the median value. In other words,
\[
A=\sum_{x\in S} |x\rangle \langle x|, \qquad S=\arg \min_{\substack{S\subseteq \{0,1\}^n \\ |S|=2^{n-1}\\ }}\; \;
\sum_{x\in S} \langle x| \calU_{\bs{\beta}}(|+\rangle \langle +|^{\otimes n})|x\rangle.
\]
By construction, $E^*(\bs{\beta})\ge 1/2$ for any circuit $\bs{\beta}$. Furthermore, we observed that $E^*(\bs{\beta})$ is well separated from $1/2$ for most of the circuits see Fig.~\ref{fig:plotIdeal}.  
Recall that we define a noisy version of a $k$-qubit unitary gate $\calU$ as $\calD_k \calU$, where 
\[
\calD_k(\rho)=(1-\epsilon) \rho +  \frac{\epsilon I}{2^k} \trace{(\rho)}
\]
is the depolarizing channel on $k$ qubits. Noise was added after all gates including the identity gates. In this case the total simulation overhead $\gamma_{\bs{\beta}}$ depends only on the number of qubits and the circuit depth, namely 
\[
\gamma_{\bs{\beta}}=\left[ \frac{1+\epsilon/2}{1-\epsilon} \right]^{nd/2}
\cdot \left[ \frac{1+7\epsilon/8}{1-\epsilon} \right]^{nd/4}.
\]
Consider a fixed ideal circuit $\calU_{\bs{\beta}}$ and let $P_{\bs{\beta}}(\bs{\alpha})$, $\calO_{\bs{\alpha}}$ be the random ensemble of noisy circuits obtained from $\calU_{\bs{\beta}}$ by inserting random Pauli operators and adding noise as described in the  main text. Instead of using the estimate Eq.~(10) of the main text for the ideal output probability $E^*(\bs{\beta})$ we opted for a slightly optimized  estimate. It is defined by dividing the total budget of $M$ runs into $K$ groups such that the $j$-th group contains $M_j$ runs
\[
M=\sum_{j=1}^K M_j.
\]
Define a random variable
\begin{equation}
\label{ppr5}
\hat{E}(\bs{\beta})\equiv    \gamma_{\bs{\beta}}   K^{-1} \sum_{j=1}^K
\sigma_{\bs{\beta}} (\bs{\alpha}^j) \frac1{M_j} \sum_{a=1}^{M_j} \langle x^a_j|A|x^a_j\rangle,
\end{equation}
where $\bs{\alpha}^1,\ldots,\bs{\alpha}^K$ are independent samples drawn from the distribution $P_{\bs{\beta}}(\bs{\alpha})$ and $x^a_j\in \{0,1\}^n$ are  readout strings obtained by measuring each qubit of the final state $\calO_{\bs{\alpha}^j}(\rho_{in})$ in the $Z$-basis. We prepare a fresh copy of the final state to generate each string $x^a_j$. Thus computing $\hat{E}(\bs{\beta})$ requires  $M$ runs of the noisy circuits with each run producing a single readout string. One can easily check that $\hat{E}(\bs{\beta})$ is an unbiased estimator of $E^*(\bs{\beta})$ for any choice of $\{M_j\}$. Our goal is to choose $\{M_j\}$ that minimize the variance of $\hat{E}(\bs{\beta})$ for a fixed $M$. One can easily check that the optimal choice is
\[
M_j \approx \frac{M \sigma_j}{\sum_{i=1}^K \sigma_i}
\]
where $\sigma_j^2 =E(\bs{\alpha}^j)-E(\bs{\alpha}^j)^2$. In order to choose optimal values of $M_j$, one has to run each circuit $\bs{\alpha}^j$ at least a few times, which gives a rough  estimate of 
$E(\bs{\alpha}^j)$ and thus $\sigma_j$. Numerical simulations were performed for the following parameters:
\[
\begin{array}{c|c}
\mbox{number of qubits} & n=6 \\
\hline
\mbox{circuit depth} & d=20 \\
\hline
\mbox{error rate} & \epsilon=0.01 \\
\hline
\mbox{total number of runs} & M=4,000 \\
\hline
\mbox{simulation overhead} & \gamma_{\bs{\beta}}\approx 4.3 \\
\end{array}
\]
Our results are presented on the left panel of Figure~\ref{fig:quasi}. For each of  $\approx 500$ ideal circuits $\bbeta$ generated at random we computed a simulation precision  $\delta(\bbeta)\equiv |\hat{E}(\bs{\beta}) - E^*(\bs{\beta})|$ where $\hat{E}(\bs{\beta})$ is the estimate defined in Eq.~(\ref{ppr5}). The plot on Figure~\ref{fig:quasi}, left, shows distribution of  the ideal circuits $\bbeta$ according to their simulation precision $\delta(\bbeta)$. The median value of $\delta(\bbeta)$ is approximately $0.05$. This is consistent with the estimate
\[
\delta(\bbeta)\approx \frac{\gamma_{\bbeta}}{\sqrt{M}} =\frac{4.3}{\sqrt{4000}} \approx 0.07.
\]
We also computed a simulation precision $\delta_0(\bbeta)$  that one would obtain  by running the circuit $\bbeta$ directly on a noisy device without error cancellation, see the right panel of Figure~\ref{fig:quasi}. It is defined as $\delta_0(\bbeta)\equiv |E(\bs{\beta}) - E^*(\bs{\beta})|$. For each circuit the output  probability $E(\bs{\beta})$ was estimated using $M=4,000$ circuit runs.  Thus the simulations presented on the left and the right panels of Figure~\ref{fig:quasi} have access to exactly the same resources. The median value of  $\delta_0(\bbeta)$ is approximately $0.15$. We conclude that error cancellation significantly improves  the simulation precision. 
\subsection{VI Quasi-probability representation for amplitude-damping noise}
A  more interesting example is the noise described by the amplitude-damping channel $\calA(\rho)=A_0 \rho A_0^\dag + A_1 \rho A_1^\dag$,  where
\[
A_0=\left[ \begin{array}{cc}
1 & 0 \\ 0 & (1-\epsilon)^{1/2} \\ \end{array} \right] \quad \mbox{and} \quad
A_1=\epsilon^{1/2} \left[ \begin{array}{cc}
0 & 1 \\ 0 & 0 \\ \end{array} \right].
\]
A noisy version of a $k$-qubit unitary  gate $\calU$ is defined as  $\calA^{\otimes k} \calU$. In contrast to the previous example, noisy unitary gates $\calA^{\otimes k} \calU$ alone cannot simulate any ideal unitary gate. Indeed, assume the contrary. Suppose $\calU_\beta$ is a single-qubit gate that has a  QPR Eq.~(\ref{local}) with $\calO_\alpha = \calA \calV_\alpha$ for some  unitary maps $\calV_\alpha$. Rewrite Eq.~(\ref{local}) as
\[
\calA^{-1} = \gamma_\beta \sum_{\alpha=1}^p P_\beta(\alpha) \sigma_\beta(\alpha) \calV_\alpha \calU_\beta^{-1}.
\]
Since the maps  $\calV_\alpha \calU_\beta^{-1}$ are unital, we infer  that $\calA^{-1}$ and $\calA$ are unital which is false. Thus Eq.~(\ref{local}) has no solutions.

To overcome this problem we shall extend the noisy basis by adding state preparations. Also we shall employ non-product QPRs. Given a single-qubit state $|\psi\rangle$, define a state preparation map 
\begin{equation}
\label{state_prep}
\calP_{|\psi\rangle}(\rho)= \trace{(\rho)}\cdot |\psi\rangle\langle \psi|.
\end{equation}
Let  $\calS(\rho)=S\rho S^{-1}$ be the $S$-gate considered as a TPCP map. Define a noisy basis $\Omega$ that includes noisy state preparations  $\calA \calP_{|\psi\rangle}$ with $|\psi\rangle=|+\rangle,|-\rangle,|0\rangle,|1\rangle$, noisy single-qubit gates $\calA \calU_\beta$, $\calA \calS^{\pm 1} \calU_\beta$ for each  ideal single-qubit gate $\calU_\beta\in \Gamma$, and noisy two-qubit gates
\[ 
\calA_c \calA_t  \calS^{y}_c \calS_t^z   \calU_{\cnot}, \quad y,z\in \{0,\pm 1\}
\]
where $c,t$ are the control and the target qubits of a CNOT gate $ \calU_{\cnot}\in \Gamma$. Here the subscripts indicate qubits acted upon by each map. We claim that this noisy basis $\Omega$ simulates any ideal Clifford+$T$ circuit $\bbeta$  with the overhead
\begin{equation}
\label{overhead1}
\gamma_{\bbeta} \le  \gamma^{L_1+2L_2}, \quad \gamma\equiv \frac{1+\epsilon}{1-\epsilon},
\end{equation}
where $L_k$  is the number of $k$-qubit gates in $\bbeta$. The corresponding QPR Eq.~(\ref{ppr1}) preserves the circuit depth, although it does not have a simple product form as above.

Indeed, consider a single-qubit gate $\calU_\beta\in \Gamma$. Let us look for a solution of Eq.~(\ref{LP1}) with $p=4$ and
\[
\calO_1=\calA\calU_\beta, \; \calO_2=\calA \calS \calU_\beta,
\;  \calO_3=\calA \calS^{-1} \calU_\beta,
\; \calO_4= \calP_{|0\rangle}.
\]
Note that $\calP_{|0\rangle} \in \Omega$ since $\calA\calP_{|0\rangle}=\calP_{|0\rangle}$. Furthermore, since $\calP_{|0\rangle}=\calA\calP_{|0\rangle}\calU_\beta$, one can rewrite Eq.~(\ref{LP1}) as 
\begin{equation}
\label{Ainv}
\calA^{-1} =\eta_1 \calI + \eta_2 \calS + \eta_3 \calS^{-1} +\eta_4 \calP_{|0\rangle}.
\end{equation}
One can easily check that the optimal solution minimizing $\sum_\alpha |\eta_\alpha|$ is 
\[
\eta_1=\frac1{\sqrt{1-\epsilon}},
\;
\eta_2=\eta_3=\frac{1-\sqrt{1-\epsilon}}{2(1-\epsilon)},
\;
\eta_4=-\frac{\epsilon}{1-\epsilon}.
\]
Therefore, $\Omega$ simulates $\calU_\beta$ with the overhead $\sum_\alpha |\eta_\alpha|=\gamma$, where $\gamma$ is defined in Eq.~(\ref{overhead1}).

Next consider the CNOT gate $\calU_{\cnot}\in \Gamma$.  Consider a decomposition of $\calA_c^{-1} \calA_t^{-1}$ obtained by applying Eq.~(\ref{Ainv}) twice. Multiplying this decomposition on the right by $\calU_{\cnot}$ and on the left by $\calA_c\calA_t$ one  obtains
\begin{equation}
\label{cnotQPR}
\calU_{\cnot}=\sum_{\alpha} \eta_\alpha \calO_{\alpha}' \calO_{\alpha},
\quad \sum_{\alpha} | \eta_{\alpha}|= \gamma^2
\end{equation}
where $\calO_{\alpha}= \calA_c \calA_t  \calS^{y}_c \calS_t^z   \calU_{\cnot}\in \Omega$ is a valid noisy operation and $\calO_{\alpha}'$ is either identity or a state preparation map $\calP_{|0\rangle}$ applied to the  control and/or target qubits. Here we noted that $\calA \calP_{|0\rangle} = \calP_{|0\rangle} \calA$. Although $\calO_{\alpha}' \calO_{\alpha}$ might not be a valid noisy operation from $\Omega$, we may merge $\calO_{\alpha}'$ with the next gate applied after the CNOT. Indeed, by assumption, each CNOT in the ideal circuit  is followed by some single-qubit gates $\calU_c$ and $\calU_t$ applied to the control and the target qubits. The gates $I,S,T$ can be absorbed into $\calP_{|0\rangle}$ since they act trivially on the state $|0\rangle$. The only non-trivial case is when $\calP_{|0\rangle}$ is merged with the Hadamard gate. In this case the latter is replaced by the state preparation $\calP_{|+\rangle}$. 

Since $\calP_{|+\rangle}$ can now appear in the ideal circuit, we must be able to use noisy operations from $\Omega$ to simulate $\calP_{|+\rangle}$. Let us look for a solution of Eq.~(\ref{LP1}) with $\calU_\beta \equiv \calP_{|+\rangle}$ in the form
\begin{equation}
\label{qprX}
\calP_{|+\rangle} = \eta_1 \calA \calP_{|+\rangle} + \eta_2 \calA \calP_{|-\rangle} +\eta_3  \calA \calP_{|1\rangle}.
\end{equation}
Note that the righthand side of Eq.~(\ref{qprX}) contains only noisy operations from $\Omega$. One can rewrite Eq.~(\ref{qprX}) as
\[
|+\rangle \langle +| = \eta_1 \calA(|+\rangle\langle +|) + \eta_2 \calA(|-\rangle\langle -|) + \eta_3 \calA(|1\rangle \langle 1).
\]
The optimal solution minimizing $\sum_\alpha |\eta_\alpha|$ is 
\[
\eta_{1,2}=\pm \frac12\left( \frac1{\sqrt{1-\epsilon}} \pm \frac{1-2\epsilon}{1-\epsilon} \right),
\quad
\eta_3=\frac{\epsilon}{1-\epsilon}.
\]
Therefore $\Omega$ simulates the ideal state preparation $\calP_{|+\rangle}$ with the overhead $\gamma'=\sum_\alpha |\eta_\alpha| \le \gamma$, where $\gamma$ is defined in Eq.~(\ref{overhead1}). 

A QPR of the ideal circuit $\bbeta$ with the overhead Eq.~(\ref{overhead1}) is constructed in two steps. First, one applies the decomposition Eq.~(\ref{cnotQPR}) to each CNOT of $\bbeta$ and merges state preparation maps $\calP_{|0\rangle}$ that appear in $\calO_{\alpha}'$ (if any) with the single-qubit gates of $\bbeta$  following the CNOT. Now all CNOT gates are replaced by noisy gates from $\Omega$. The rest of the circuit consists of single-qubit gates $\calU_\beta \in \Gamma$ and state preparations $\calP_{|+\rangle}$. At the second step, each of these ideal operations is replaced by  its QPR constructed above. Note that each CNOT contributes $\gamma^2$ to the total overhead $\gamma_{\bbeta}$, see Eq.~(\ref{cnotQPR}). Each single-qubit gate $\calU_\beta\in \Gamma$ or a state preparation $\calP_{|+\rangle}$ contributes at most $\gamma$ to the total overhead. This proves Eq.~(\ref{overhead1}).
\end{appendix}

\begin{thebibliography}{25}%
\makeatletter
\providecommand \@ifxundefined [1]{%
 \@ifx{#1\undefined}
}%
\providecommand \@ifnum [1]{%
 \ifnum #1\expandafter \@firstoftwo
 \else \expandafter \@secondoftwo
 \fi
}%
\providecommand \@ifx [1]{%
 \ifx #1\expandafter \@firstoftwo
 \else \expandafter \@secondoftwo
 \fi
}%
\providecommand \natexlab [1]{#1}%
\providecommand \enquote  [1]{``#1''}%
\providecommand \bibnamefont  [1]{#1}%
\providecommand \bibfnamefont [1]{#1}%
\providecommand \citenamefont [1]{#1}%
\providecommand \href@noop [0]{\@secondoftwo}%
\providecommand \href [0]{\begingroup \@sanitize@url \@href}%
\providecommand \@href[1]{\@@startlink{#1}\@@href}%
\providecommand \@@href[1]{\endgroup#1\@@endlink}%
\providecommand \@sanitize@url [0]{\catcode `\\12\catcode `\$12\catcode
  `\&12\catcode `\#12\catcode `\^12\catcode `\_12\catcode `\%12\relax}%
\providecommand \@@startlink[1]{}%
\providecommand \@@endlink[0]{}%
\providecommand \url  [0]{\begingroup\@sanitize@url \@url }%
\providecommand \@url [1]{\endgroup\@href {#1}{\urlprefix }}%
\providecommand \urlprefix  [0]{URL }%
\providecommand \Eprint [0]{\href }%
\providecommand \doibase [0]{http://dx.doi.org/}%
\providecommand \selectlanguage [0]{\@gobble}%
\providecommand \bibinfo  [0]{\@secondoftwo}%
\providecommand \bibfield  [0]{\@secondoftwo}%
\providecommand \translation [1]{[#1]}%
\providecommand \BibitemOpen [0]{}%
\providecommand \bibitemStop [0]{}%
\providecommand \bibitemNoStop [0]{.\EOS\space}%
\providecommand \EOS [0]{\spacefactor3000\relax}%
\providecommand \BibitemShut  [1]{\csname bibitem#1\endcsname}%
\let\auto@bib@innerbib\@empty
%</preamble>
\bibitem [{\citenamefont {Unruh}(1995)}]{unruh1995maintaining}%
  \BibitemOpen
  \bibfield  {author} {\bibinfo {author} {\bibfnamefont {W.~G.}\ \bibnamefont
  {Unruh}},\ }\href@noop {} {\bibfield  {journal} {\bibinfo  {journal}
  {Physical Review A}\ }\textbf {\bibinfo {volume} {51}},\ \bibinfo {pages}
  {992} (\bibinfo {year} {1995})}\BibitemShut {NoStop}%
\bibitem [{\citenamefont {Shor}(1995)}]{shor1995scheme}%
  \BibitemOpen
  \bibfield  {author} {\bibinfo {author} {\bibfnamefont {P.~W.}\ \bibnamefont
  {Shor}},\ }\href@noop {} {\bibfield  {journal} {\bibinfo  {journal} {Physical
  review A}\ }\textbf {\bibinfo {volume} {52}},\ \bibinfo {pages} {R2493}
  (\bibinfo {year} {1995})}\BibitemShut {NoStop}%
\bibitem [{\citenamefont {Steane}(1996)}]{steane1996error}%
  \BibitemOpen
  \bibfield  {author} {\bibinfo {author} {\bibfnamefont {A.~M.}\ \bibnamefont
  {Steane}},\ }\href@noop {} {\bibfield  {journal} {\bibinfo  {journal}
  {Physical Review Letters}\ }\textbf {\bibinfo {volume} {77}},\ \bibinfo
  {pages} {793} (\bibinfo {year} {1996})}\BibitemShut {NoStop}%
\bibitem [{\citenamefont {Calderbank}\ and\ \citenamefont
  {Shor}(1996)}]{calderbank1996good}%
  \BibitemOpen
  \bibfield  {author} {\bibinfo {author} {\bibfnamefont {A.~R.}\ \bibnamefont
  {Calderbank}}\ and\ \bibinfo {author} {\bibfnamefont {P.~W.}\ \bibnamefont
  {Shor}},\ }\href@noop {} {\bibfield  {journal} {\bibinfo  {journal} {Physical
  Review A}\ }\textbf {\bibinfo {volume} {54}},\ \bibinfo {pages} {1098}
  (\bibinfo {year} {1996})}\BibitemShut {NoStop}%
\bibitem [{\citenamefont {Aharonov}\ and\ \citenamefont
  {Ben-Or}(1997)}]{aharonov1997fault}%
  \BibitemOpen
  \bibfield  {author} {\bibinfo {author} {\bibfnamefont {D.}~\bibnamefont
  {Aharonov}}\ and\ \bibinfo {author} {\bibfnamefont {M.}~\bibnamefont
  {Ben-Or}},\ }in\ \href@noop {} {\emph {\bibinfo {booktitle} {Proceedings of
  the twenty-ninth annual ACM symposium on Theory of computing}}}\ (\bibinfo
  {organization} {ACM},\ \bibinfo {year} {1997})\ pp.\ \bibinfo {pages}
  {176--188}\BibitemShut {NoStop}%
\bibitem [{\citenamefont {Kitaev}(1997)}]{kitaev1997quantum}%
  \BibitemOpen
  \bibfield  {author} {\bibinfo {author} {\bibfnamefont {A.~Y.}\ \bibnamefont
  {Kitaev}},\ }\href@noop {} {\bibfield  {journal} {\bibinfo  {journal}
  {Russian Mathematical Surveys}\ }\textbf {\bibinfo {volume} {52}},\ \bibinfo
  {pages} {1191} (\bibinfo {year} {1997})}\BibitemShut {NoStop}%
\bibitem [{\citenamefont {Fowler}\ \emph {et~al.}(2012)\citenamefont {Fowler},
  \citenamefont {Mariantoni}, \citenamefont {Martinis},\ and\ \citenamefont
  {Cleland}}]{fowler2012surface}%
  \BibitemOpen
  \bibfield  {author} {\bibinfo {author} {\bibfnamefont {A.~G.}\ \bibnamefont
  {Fowler}}, \bibinfo {author} {\bibfnamefont {M.}~\bibnamefont {Mariantoni}},
  \bibinfo {author} {\bibfnamefont {J.~M.}\ \bibnamefont {Martinis}}, \ and\
  \bibinfo {author} {\bibfnamefont {A.~N.}\ \bibnamefont {Cleland}},\
  }\href@noop {} {\bibfield  {journal} {\bibinfo  {journal} {Physical Review
  A}\ }\textbf {\bibinfo {volume} {86}},\ \bibinfo {pages} {032324} (\bibinfo
  {year} {2012})}\BibitemShut {NoStop}%
\bibitem [{\citenamefont {Jones}\ \emph {et~al.}(2012)\citenamefont {Jones},
  \citenamefont {Van~Meter}, \citenamefont {Fowler}, \citenamefont {McMahon},
  \citenamefont {Kim}, \citenamefont {Ladd},\ and\ \citenamefont
  {Yamamoto}}]{jones2012layered}%
  \BibitemOpen
  \bibfield  {author} {\bibinfo {author} {\bibfnamefont {N.~C.}\ \bibnamefont
  {Jones}}, \bibinfo {author} {\bibfnamefont {R.}~\bibnamefont {Van~Meter}},
  \bibinfo {author} {\bibfnamefont {A.~G.}\ \bibnamefont {Fowler}}, \bibinfo
  {author} {\bibfnamefont {P.~L.}\ \bibnamefont {McMahon}}, \bibinfo {author}
  {\bibfnamefont {J.}~\bibnamefont {Kim}}, \bibinfo {author} {\bibfnamefont
  {T.~D.}\ \bibnamefont {Ladd}}, \ and\ \bibinfo {author} {\bibfnamefont
  {Y.}~\bibnamefont {Yamamoto}},\ }\href@noop {} {\bibfield  {journal}
  {\bibinfo  {journal} {Physical Review X}\ }\textbf {\bibinfo {volume} {2}},\
  \bibinfo {pages} {031007} (\bibinfo {year} {2012})}\BibitemShut {NoStop}%
\bibitem [{\citenamefont {Devitt}\ \emph {et~al.}(2013)\citenamefont {Devitt},
  \citenamefont {Stephens}, \citenamefont {Munro},\ and\ \citenamefont
  {Nemoto}}]{devitt2013requirements}%
  \BibitemOpen
  \bibfield  {author} {\bibinfo {author} {\bibfnamefont {S.~J.}\ \bibnamefont
  {Devitt}}, \bibinfo {author} {\bibfnamefont {A.~M.}\ \bibnamefont
  {Stephens}}, \bibinfo {author} {\bibfnamefont {W.~J.}\ \bibnamefont {Munro}},
  \ and\ \bibinfo {author} {\bibfnamefont {K.}~\bibnamefont {Nemoto}},\
  }\href@noop {} {\bibfield  {journal} {\bibinfo  {journal} {Nature
  communications}\ }\textbf {\bibinfo {volume} {4}} (\bibinfo {year}
  {2013})}\BibitemShut {NoStop}%
\bibitem [{\citenamefont {Peruzzo}\ \emph {et~al.}(2014)\citenamefont
  {Peruzzo}, \citenamefont {McClean}, \citenamefont {Shadbolt}, \citenamefont
  {Yung}, \citenamefont {Zhou}, \citenamefont {Love}, \citenamefont
  {Aspuru-Guzik},\ and\ \citenamefont {O?Brien}}]{peruzzo2014variational}%
  \BibitemOpen
  \bibfield  {author} {\bibinfo {author} {\bibfnamefont {A.}~\bibnamefont
  {Peruzzo}}, \bibinfo {author} {\bibfnamefont {J.}~\bibnamefont {McClean}},
  \bibinfo {author} {\bibfnamefont {P.}~\bibnamefont {Shadbolt}}, \bibinfo
  {author} {\bibfnamefont {M.-H.}\ \bibnamefont {Yung}}, \bibinfo {author}
  {\bibfnamefont {X.-Q.}\ \bibnamefont {Zhou}}, \bibinfo {author}
  {\bibfnamefont {P.~J.}\ \bibnamefont {Love}}, \bibinfo {author}
  {\bibfnamefont {A.}~\bibnamefont {Aspuru-Guzik}}, \ and\ \bibinfo {author}
  {\bibfnamefont {J.~L.}\ \bibnamefont {O?Brien}},\ }\href@noop {} {\bibfield
  {journal} {\bibinfo  {journal} {Nature communications}\ }\textbf {\bibinfo
  {volume} {5}} (\bibinfo {year} {2014})}\BibitemShut {NoStop}%
\bibitem [{\citenamefont {McClean}\ \emph {et~al.}(2015)\citenamefont
  {McClean}, \citenamefont {Romero}, \citenamefont {Babbush},\ and\
  \citenamefont {Aspuru-Guzik}}]{mcclean2015theory}%
  \BibitemOpen
  \bibfield  {author} {\bibinfo {author} {\bibfnamefont {J.~R.}\ \bibnamefont
  {McClean}}, \bibinfo {author} {\bibfnamefont {J.}~\bibnamefont {Romero}},
  \bibinfo {author} {\bibfnamefont {R.}~\bibnamefont {Babbush}}, \ and\
  \bibinfo {author} {\bibfnamefont {A.}~\bibnamefont {Aspuru-Guzik}},\
  }\href@noop {} {\bibfield  {journal} {\bibinfo  {journal} {arXiv preprint
  arXiv:1509.04279}\ } (\bibinfo {year} {2015})}\BibitemShut {NoStop}%
\bibitem [{\citenamefont {Wecker}\ \emph {et~al.}(2015)\citenamefont {Wecker},
  \citenamefont {Hastings},\ and\ \citenamefont {Troyer}}]{wecker2015progress}%
  \BibitemOpen
  \bibfield  {author} {\bibinfo {author} {\bibfnamefont {D.}~\bibnamefont
  {Wecker}}, \bibinfo {author} {\bibfnamefont {M.~B.}\ \bibnamefont
  {Hastings}}, \ and\ \bibinfo {author} {\bibfnamefont {M.}~\bibnamefont
  {Troyer}},\ }\href@noop {} {\bibfield  {journal} {\bibinfo  {journal}
  {Physical Review A}\ }\textbf {\bibinfo {volume} {92}},\ \bibinfo {pages}
  {042303} (\bibinfo {year} {2015})}\BibitemShut {NoStop}%
\bibitem [{\citenamefont {Farhi}\ \emph {et~al.}(2014)\citenamefont {Farhi},
  \citenamefont {Goldstone},\ and\ \citenamefont {Gutmann}}]{farhi2014quantum}%
  \BibitemOpen
  \bibfield  {author} {\bibinfo {author} {\bibfnamefont {E.}~\bibnamefont
  {Farhi}}, \bibinfo {author} {\bibfnamefont {J.}~\bibnamefont {Goldstone}}, \
  and\ \bibinfo {author} {\bibfnamefont {S.}~\bibnamefont {Gutmann}},\
  }\href@noop {} {\bibfield  {journal} {\bibinfo  {journal} {arXiv preprint
  arXiv:1411.4028}\ } (\bibinfo {year} {2014})}\BibitemShut {NoStop}%
\bibitem [{\citenamefont {Bauer}\ \emph {et~al.}(2015)\citenamefont {Bauer},
  \citenamefont {Wecker}, \citenamefont {Millis}, \citenamefont {Hastings},\
  and\ \citenamefont {Troyer}}]{bauer2015hybrid}%
  \BibitemOpen
  \bibfield  {author} {\bibinfo {author} {\bibfnamefont {B.}~\bibnamefont
  {Bauer}}, \bibinfo {author} {\bibfnamefont {D.}~\bibnamefont {Wecker}},
  \bibinfo {author} {\bibfnamefont {A.~J.}\ \bibnamefont {Millis}}, \bibinfo
  {author} {\bibfnamefont {M.~B.}\ \bibnamefont {Hastings}}, \ and\ \bibinfo
  {author} {\bibfnamefont {M.}~\bibnamefont {Troyer}},\ }\href@noop {}
  {\bibfield  {journal} {\bibinfo  {journal} {arXiv preprint arXiv:1510.03859}\
  } (\bibinfo {year} {2015})}\BibitemShut {NoStop}%
\bibitem [{\citenamefont {Richardson}\ and\ \citenamefont
  {Gaunt}(1927)}]{richardson1927deferred}%
  \BibitemOpen
  \bibfield  {author} {\bibinfo {author} {\bibfnamefont {L.~F.}\ \bibnamefont
  {Richardson}}\ and\ \bibinfo {author} {\bibfnamefont {J.~A.}\ \bibnamefont
  {Gaunt}},\ }\href@noop {} {\bibfield  {journal} {\bibinfo  {journal}
  {Philosophical Transactions of the Royal Society of London. Series A,
  containing papers of a mathematical or physical character}\ }\textbf
  {\bibinfo {volume} {226}},\ \bibinfo {pages} {299} (\bibinfo {year}
  {1927})}\BibitemShut {NoStop}%
\bibitem [{\citenamefont {Sidi}(2003)}]{sidi2003practical}%
  \BibitemOpen
  \bibfield  {author} {\bibinfo {author} {\bibfnamefont {A.}~\bibnamefont
  {Sidi}},\ }\href@noop {} {\enquote {\bibinfo {title} {Practical extrapolation
  methods: Theory and applications, volume 10 of cambridge monographs on
  applied and computational mathematics},}\ } (\bibinfo {year}
  {2003})\BibitemShut {NoStop}%
\bibitem [{\citenamefont {Pashayan}\ \emph {et~al.}(2015)\citenamefont
  {Pashayan}, \citenamefont {Wallman},\ and\ \citenamefont
  {Bartlett}}]{Pashayan2015}%
  \BibitemOpen
  \bibfield  {author} {\bibinfo {author} {\bibfnamefont {H.}~\bibnamefont
  {Pashayan}}, \bibinfo {author} {\bibfnamefont {J.~J.}\ \bibnamefont
  {Wallman}}, \ and\ \bibinfo {author} {\bibfnamefont {S.~D.}\ \bibnamefont
  {Bartlett}},\ }\href@noop {} {\bibfield  {journal} {\bibinfo  {journal}
  {Physical review letters}\ }\textbf {\bibinfo {volume} {115}},\ \bibinfo
  {pages} {070501} (\bibinfo {year} {2015})}\BibitemShut {NoStop}%
\bibitem [{\citenamefont {Delfosse}\ \emph {et~al.}(2015)\citenamefont
  {Delfosse}, \citenamefont {Guerin}, \citenamefont {Bian},\ and\ \citenamefont
  {Raussendorf}}]{Delfosse2015}%
  \BibitemOpen
  \bibfield  {author} {\bibinfo {author} {\bibfnamefont {N.}~\bibnamefont
  {Delfosse}}, \bibinfo {author} {\bibfnamefont {P.~A.}\ \bibnamefont
  {Guerin}}, \bibinfo {author} {\bibfnamefont {J.}~\bibnamefont {Bian}}, \ and\
  \bibinfo {author} {\bibfnamefont {R.}~\bibnamefont {Raussendorf}},\
  }\href@noop {} {\bibfield  {journal} {\bibinfo  {journal} {Physical Review
  X}\ }\textbf {\bibinfo {volume} {5}},\ \bibinfo {pages} {021003} (\bibinfo
  {year} {2015})}\BibitemShut {NoStop}%
\bibitem [{\citenamefont {Mohseni}\ \emph {et~al.}(2008)\citenamefont
  {Mohseni}, \citenamefont {Rezakhani},\ and\ \citenamefont
  {Lidar}}]{mohseni2008quantum}%
  \BibitemOpen
  \bibfield  {author} {\bibinfo {author} {\bibfnamefont {M.}~\bibnamefont
  {Mohseni}}, \bibinfo {author} {\bibfnamefont {A.}~\bibnamefont {Rezakhani}},
  \ and\ \bibinfo {author} {\bibfnamefont {D.}~\bibnamefont {Lidar}},\
  }\href@noop {} {\bibfield  {journal} {\bibinfo  {journal} {Phys. Rev. A}\
  }\textbf {\bibinfo {volume} {77}},\ \bibinfo {pages} {032322} (\bibinfo
  {year} {2008})}\BibitemShut {NoStop}%
\bibitem [{\citenamefont {Silva}\ \emph {et~al.}(2008)\citenamefont {Silva},
  \citenamefont {Magesan}, \citenamefont {Kribs},\ and\ \citenamefont
  {Emerson}}]{Silva2008}%
  \BibitemOpen
  \bibfield  {author} {\bibinfo {author} {\bibfnamefont {M.}~\bibnamefont
  {Silva}}, \bibinfo {author} {\bibfnamefont {E.}~\bibnamefont {Magesan}},
  \bibinfo {author} {\bibfnamefont {D.~W.}\ \bibnamefont {Kribs}}, \ and\
  \bibinfo {author} {\bibfnamefont {J.}~\bibnamefont {Emerson}},\ }\href@noop
  {} {\bibfield  {journal} {\bibinfo  {journal} {Physical Review A}\ }\textbf
  {\bibinfo {volume} {78}},\ \bibinfo {pages} {012347} (\bibinfo {year}
  {2008})}\BibitemShut {NoStop}%
\bibitem [{\citenamefont {Magesan}\ \emph {et~al.}(2012)\citenamefont
  {Magesan}, \citenamefont {Gambetta},\ and\ \citenamefont
  {Emerson}}]{Magesan2012}%
  \BibitemOpen
  \bibfield  {author} {\bibinfo {author} {\bibfnamefont {E.}~\bibnamefont
  {Magesan}}, \bibinfo {author} {\bibfnamefont {J.~M.}\ \bibnamefont
  {Gambetta}}, \ and\ \bibinfo {author} {\bibfnamefont {J.}~\bibnamefont
  {Emerson}},\ }\href@noop {} {\bibfield  {journal} {\bibinfo  {journal}
  {Physical Review A}\ }\textbf {\bibinfo {volume} {85}},\ \bibinfo {pages}
  {042311} (\bibinfo {year} {2012})}\BibitemShut {NoStop}%
\bibitem [{\citenamefont {Li}\ and\ \citenamefont
  {Benjamin}(2016)}]{li2016efficient}%
  \BibitemOpen
  \bibfield  {author} {\bibinfo {author} {\bibfnamefont {Y.}~\bibnamefont
  {Li}}\ and\ \bibinfo {author} {\bibfnamefont {S.~C.}\ \bibnamefont
  {Benjamin}},\ }\href@noop {} {\bibfield  {journal} {\bibinfo  {journal}
  {arXiv preprint arXiv:1611.09301}\ } (\bibinfo {year} {2016})}\BibitemShut
  {NoStop}%
\bibitem [{\citenamefont {Lindblad}(1976)}]{lindblad1976generators}%
  \BibitemOpen
  \bibfield  {author} {\bibinfo {author} {\bibfnamefont {G.}~\bibnamefont
  {Lindblad}},\ }\href@noop {} {\bibfield  {journal} {\bibinfo  {journal}
  {Communications in Mathematical Physics}\ }\textbf {\bibinfo {volume} {48}},\
  \bibinfo {pages} {119} (\bibinfo {year} {1976})}\BibitemShut {NoStop}%
\bibitem [{\citenamefont {Breuer}\ and\ \citenamefont
  {Petruccione}(2002)}]{breuer2002theory}%
  \BibitemOpen
  \bibfield  {author} {\bibinfo {author} {\bibfnamefont {H.-P.}\ \bibnamefont
  {Breuer}}\ and\ \bibinfo {author} {\bibfnamefont {F.}~\bibnamefont
  {Petruccione}},\ }\href@noop {} {\emph {\bibinfo {title} {The theory of open
  quantum systems}}}\ (\bibinfo  {publisher} {Oxford University Press on
  Demand},\ \bibinfo {year} {2002})\BibitemShut {NoStop}%
\bibitem [{\citenamefont {Schneider}(1975)}]{schneider1975vereinfachte}%
  \BibitemOpen
  \bibfield  {author} {\bibinfo {author} {\bibfnamefont {C.}~\bibnamefont
  {Schneider}},\ }\href@noop {} {\bibfield  {journal} {\bibinfo  {journal}
  {Numerische Mathematik}\ }\textbf {\bibinfo {volume} {24}},\ \bibinfo {pages}
  {177} (\bibinfo {year} {1975})}\BibitemShut {NoStop}%
\end{thebibliography}
\end{document}